\documentclass[12pt]{article}
\usepackage{latexsym}
\usepackage{amssymb}
\usepackage{amsmath}
\usepackage{graphicx}
\usepackage{enumitem}
\usepackage{slashed}
\usepackage{hyperref}

\newcommand\mathC{\mkern1mu\raise2.2pt\hbox{$\scriptscriptstyle|$}
        {\mkern-7mu\rm C}}              

\def\be{\begin{equation}}
\def\ee{\end{equation}}
\def\bear{\begin{eqnarray}}
\def\eear{\end{eqnarray}}
\def\nn{\nonumber}

\newcommand\bra[1]{{\langle {#1}|}}
\newcommand\ket[1]{{|{#1}\rangle}}

\def\r{\rho}

\def\s{\sigma}

\def\dd{\mbox{d}}

\def\bra{\langle}
\def\ket{\rangle}

\def\l{\lambda}
\def\L{\Lambda}
\def\m{\mu}
\def\n{\nu}

\def\r{\rho}

\newcommand{\ti}[1]{\tilde{#1}}

\newcommand{\sm}[1]{\mbox{\scriptsize #1}}
\newcommand{\tn}[1]{\mbox{\tiny #1}}
\renewcommand{\@}[1]{\sqrt{#1}}
\renewcommand{\le}[1]{\label{#1}\end{eqnarray}}
\newcommand{\bea}{\begin{eqnarray}}
\newcommand{\eea}{\end{eqnarray}}

\newcommand{\eq}[1]{(\ref{#1})}
\def\nn{\nonumber\\}

\def\ffract#1#2{\raise .35 em\hbox{$\scriptstyle#1$}\kern-.25em/
\kern-.2em\lower .22 em \hbox{$\scriptstyle#2$}}

\begin{document}

\pagestyle{empty}

\centerline{{\Large \bf Theoretical Equivalence and Duality}}
\vskip1cm

\begin{center}
{\large Sebastian De Haro}\\
\vskip 1truecm
{\it Trinity College, Cambridge, CB2 1TQ, United Kingdom}\\
{\it Department of History and Philosophy of Science, University of Cambridge}\\
{\it Vossius Center for History of Humanities and Sciences, University of Amsterdam}

\vskip .7truecm
{\tt sd696@cam.ac.uk}
\vskip 1cm
\today
\end{center}

\vskip 10truecm

\begin{center}
\textbf{\large \bf Abstract}
\end{center}

Theoretical equivalence and duality are two closely related notions: but their interconnection has so far not been well understood. In this paper I explicate the contribution of a recent schema for duality to discussions of theoretical equivalence. I argue that duality suggests a construal of theoretical equivalence in the physical sciences.  The construal is in terms of the isomorphism of models, as defined by the Schema. This construal entails interpretative constraints that should be useful for theoretical equivalence more generally. I illustrate the construal in various formulations of Maxwell's electromagnetic theory.

\newpage
\pagestyle{plain}

\tableofcontents

\newpage

\section{Introduction}\label{intro}

Theoretical equivalence and duality are both taken to be, roughly, a matter of ``two theories saying the same thing, in different words''. Thus the question arises how these two notions are related: and, in particular, whether discussions of duality bear on discussions of theoretical equivalence. This paper aims to address this question. I will argue that a recently proposed Schema for physical theories and duality (De Haro (2016) and De Haro and Butterfield (2017)) leads to a proposal for theoretical equivalence that, furthermore, suggests interpretative constraints on the use of theoretical equivalence in the physical sciences (Section \ref{schemadte}). The proposal will be illustrated in a well-known case in the literature of theoretical equivalence, namely the equivalence between various formulations of Maxwell's electromagnetic theory (Section \ref{Maxem}).\\

Theoretical equivalence is an old and venerable topic in the philosophy of science, which goes back to the logical positivists: and it finds its roots even earlier, for example in the philosophy of spacetime (viz.~in the Leibniz-Clarke correspondence, and in Poincar\'e's conventionalism). Two influential accounts are Quine (1975) and Glymour (1970). The recent discussion of theoretical equivalence, with which I will engage, includes: Halvorson (2012, 2013), Glymour (2013), van Fraassen (2014), Coffey (2014), Weatherall (2015, 2016, 2016a), Barrett and Halvorson  (2016), Lutz (2017), Teh and Tsementzis (2017), Barrett (2018), and Hudetz (2018).

Dualities have come to stand centre stage in theory-construction in theoretical physics: since, at least, the discoveries of position-momentum duality and electric-magnetic duality, in quantum mechanics. But even more so, recently: in statistical physics, condensed matter physics, quantum field theory, and quantum gravity. Accordingly, the philosophical literature on duality is now flourishing.\footnote{See the recent special issue on dualities edited by Castellani and Rickles (2017). Other recent philosophical discussions of duality include De Haro (2016), Read (2016), De Haro and Butterfield (2017), and Butterfield (2018).}

Although the recent literature on dualities has of course engaged with the broad philosophical discussion of theoretical equivalence,\footnote{Earlier work relating duality and the recent literature on theoretical equivalence is in De Haro (2016:~Section 1.4), Butterfield (2018), Read and M\o ller-Nielsen (2018:~Section 4.1). See also Weatherall (2019).} it has for the most part not attempted to answer in detail the question of how duality {\it bears} on the notion of theoretical equivalence. This paper aims to fill this gap, by developing a duality-inspired criterion of theoretical equivalence, which I will dub {\it physical equivalence}. 

Recent discussions of theoretical equivalence have focussed on an important question: {\it what is the best formal account of equivalence between physical theories?} In an influential series of 
papers, Weatherall (2015, 2016, 2016a) has argued that categorical equivalence provides a good standard of theoretical equivalence for examples in physics. He also proposes an interesting category-theoretic criterion for ``when a theory has excess structure'': namely, when an appropriate functor between the two empirically equivalent theories ``forgets structure''.\footnote{For related work, see Halvorson (2016), Halvorson and Tsementzis (2015), and Barrett (2018).} He has argued that categorical equivalence gives correct verdicts in a number of important cases: Newtonian gravitation vis-\`a-vis Newton-Cartan theory, various versions of electromagnetism, Yang-Mills theory, etc. On the other hand, Barrett and Halvorson (2016:~p.~556), and Hudetz (2018:~\S2.1), have argued that categorical equivalence is ``too liberal'', in that it rules as equivalent theories that we would not expect to count as equivalent. Thus it is worth exploring other formal conceptions of theoretical equivalence in connection with Weatherall's examples, to see whether they make the same judgments.

But the search for a formal account should not overshadow the importance of {\it interpretation}. I will argue that the formulation of the formal account is itself dependent on interpretation, since theoretical equivalence usually requires a suitable ``translation''. And so, I will argue that an account of theoretical equivalence requires having an account of semantic interpretation. I will argue that the Schema's distinction between internal and external interpretations\footnote{An early version of this distinction, as internal vs.~external views on duality, is in Dieks et al.~(2015:~p.~209) and De Haro (2015:~p.~116).}  gives just such an account, and from it that one can draw interpretative lessons for theoretical equivalence more generally.

First, a warning about the Schema's use of `theory' and `model'. The Schema uses `model' for a representation of a theory. This usage {\it differs} from the normal usage of a model as a particular solution of a theory (or a particular trajectory in the space of states). What we here call a `model' is often called, in philosophy of physics, a `theory'. This usage is motivated by dualities: dualities relate different `theories', as being different formulations of `one underlying theory': and so, they suggest that we should push the usage of both `theory' and `model' ``one level up''.\footnote{Thus I will refer to the solutions simply as `solutions' or `possible worlds', rather than `models'.}
In view of this basic fact about dualities---what we first considered as distinct theories is now seen as the same theory, in two guises---we are led to allow a more general notion of theory, and of model.

Therefore, I will propose, roughly speaking, the following construals of the three crucial terms:\\
\\
(i)~~~{\it Duality:} isomorphism of models of a single theory.\\
(ii)~~{\it Weak theoretical equivalence:} isomorphism of models (of a single theory) and matching of interpretations (where `matching of interpretations' means that the domains of application used in the interpretations are isomorphic).\\
(iii)~~{\it Physical equivalence, as the duality-based account of theoretical equivalence:} sameness of the interpretations of weakly theoretically equivalent models (where `sameness of the interpretations' means the lack of a difference between the elements and relations in the domains of application of the two models).\\
\\
Thus the Schema's notion of theoretical equivalence, namely physical equivalence, is an isomorphism criterion of equivalence. Such criteria have been criticised, in the context of the semantic view of theories, because they supposedly make distinctions without a difference---and this criticism is often accompanied by repudiation of the semantic view. These criticisms have motivated the search for other criteria of theoretical equivalence.

However, Lutz (2017:~p.~335) has recently argued that the arguments against the isomorphism criterion can be blocked: and furthermore, that the recent syntax-semantics debate does really not capture any significant differences (ibid., p.~347). Likewise, Hudetz (2018:~p.~18) has introduced a new criterion, definable categorical equivalence, which proposes to find a middle-ground between the category-theoretic approach and the earlier definability-theoretic approach exemplified by Glymour (1970, 1977) and Quine (1975). 

I will add to this discussion by arguing that some of the criticisms of the isomorphism criterion have as their target only `na\"ive' notions of isomorphism, and do not impugn more sophisticated notions such as the one provided by the Schema. And so, some of the motivations recently adduced for abandoning the isomorphism criterion will be found to be wanting. \\


I do not mean to be dogmatic about the Schema's formal adequacy for dealing with all possible theories.  There may be more generally valid, or more precise, mathematical conceptions of theoretical equivalence than the one I will work out based on the Schema. But it is worth exploring how far the Schema can go, and we will see that it fares very well: for it is able to give undoubtedly reasonable judgments about the case study presented in this paper (Maxwell's electromagnetic theory) and it also gives correct judgments about previous case studies (bosonization and gauge-gravity duality). 
Further case studies are underway.

I am however uncompromising about the conceptual and methodological importance of semantic interpretation in any account of theoretical equivalence: a feature endorsed by the Schema. Errors in the formalism are generally easy to spot, while interpretative oversights often remain hidden in the background---and they do bear on judgments of theoretical equivalence. And so, I think that one important way in which the literature about dualities can contribute to general discussions of theoretical equivalence is through the philosophical analyses of the interpretation of duals that it has developed in recent years.

I will give, in Section \ref{schemadte}, an account of physical equivalence---roughly, the idea that two dual theories describe the ``same sector of reality''. My analysis will underline that a full discussion of physical equivalence requires mathematically consistent theories---so that discussions of theoretical equivalence in the context of, say, Newtonian gravitation, are at best useful toy models.

The plan of the paper is as follows. Section \ref{schemadte} gives a proposal for theoretical equivalence, based on the Schema, which it then compares with physical equivalence. Section \ref{Maxem} contains the case study, namely Maxwell's theory. Section \ref{conclusion} concludes.

\section{The Schema as a Proposal for Theoretical Equivalence}\label{schemadte}

I first present, in Section \ref{schema}, the Schema for theories and duality: from De Haro (2016:~\S1.1-\S1.2) and De Haro and Butterfield (2017:~\S2.2-\S2.3). Section \ref{nteq} contains the proposal for theoretical equivalence, based on this Schema. Section \ref{diffeq} contrasts the notions of theoretical equivalence and physical equivalence. 

\subsection{The Schema for theories and duality}\label{schema}

In this Section, I introduce the Schema for duality. I first introduce, in Section \ref{thmod}, the notions of theories and models. In Section \ref{seminter} I discuss interpretation. I then introduce, in Section \ref{0dual}, duality and the notion of {\it physical} equivalence. Finally, in Section \ref{agiso}, I say how the notion of isomorphism in the Schema differs from what in Section \ref{intro} I called `na\"ive' isomorphism accounts. 

\subsubsection{Theories and models}\label{thmod}

Recall that dualities reveal that what we thought were distinct theories are representations of a single theory: so that our talk of `theories' and `models' is pushed ``one level up''. Thus, in this Section, I discuss the notions of theory, model, and interpretation, as construed by the Schema. 

The core notion of the Schema is that of a {\it bare theory}: a physically uninterpreted, but mathematically formulated, structure with a set of rules for forming sentences, i.e.~an abstract calculus (we will refer to these rules as the `language' of the theory: see below).\footnote{The Schema's account of bare theories bears some similarities to the framework of van Fraassen (1970:~pp.~328-329, 2014:~pp.~281-282). What I will refer to as the `sentences' of the bare theory are similar to van Fraassen's `elementary statements'. One may object that theories in physics always come to us interpreted. But one should not take the conceptual analysis that I am doing here, namely dissecting a theory into its bare, uninterpreted, part and its interpretation, for the temporal or methodological process of formulating or learning new theories. Two reasons for this are as follows: (1) Although theories of physics usually come to us interpreted, it is perfectly sensible to conceptually distinguish between the formalism of a theory and its interpretation. (2) Bare theories often have {\it more than one interpretation,} and the bare theory captures the formal structure that underlies them: for example, the heat equation can be interpreted as describing the spread of heat in time in a given region, or as a diffusion equation for Brownian particles in a medium.} A bare theory could consist of a set of axioms or a set of equations. But, to be specific, the Schema considers a bare theory as a triple, $T:=\bra{\cal S},{\cal Q},{\cal D}\ket$, of a structured state-space, ${\cal S}$, a structured set of quantities, ${\cal Q}$, and a dynamics, ${\cal D}$, consistent with the relevant structure. `Structure' here refers: first, to symmetries which may act on the states and-or the quantities, e.g.~as automorphisms of the state-space, $a:{\cal S}\rightarrow{\cal S}$. And second, `structure' also refers to the set of rules for forming sentences, e.g.~for assigning values to the quantities.\footnote{These two cases of `structure' are of course not exhaustive. For example, state-spaces often come equipped with other structures: a topology, a symplectic form, etc. But we can take this in our stride---we will not need, nor is it I think meaningful, to try and specify all the relevant structure once and for all: it can be done case by case.}

As an example, consider Maxwell's electromagnetic theory in vacuum, on $\mathbb{R}^4$. This theory can be written as a triple, of which the state-space, ${\cal S}$, is the space of square-integrable, smooth 2-forms $F$ (or ``Faraday tensors''). A salient quantity in this theory (of course not the only one: for more details, see Section \ref{Maxem}) is the ``stress-energy tensor'',\footnote{The scare quotes indicate that at this stage we have not yet interpreted the theory, and so names like ``Faraday tensor'' or ``stress-energy tensor'' are only labels that facilitate that the reader may recognise the familiar notions from Maxwell's electromagnetic theory. But at this stage one might simply drop these labels: the 2-form $F$ and the symmetric tensor $t$ could indeed describe any other field satisfying the same dynamics.} $t[F]$: namely, a rank-2 smooth, symmetric, conserved tensor, quadratic in $F$: cf.~Eq.~\eq{Tmn}. (I use the lowercase $t$ here because $T$ already stands for `theory'. The tensor components of $t$ are usually written as $T_{\m\n}$). We can think of quantities ${\cal Q}$ as maps from the state-space to a field of real or complex numbers (so that ${\cal Q}$ is the dual space---in the mathematical, not the Schema's sense!---of ${\cal S}$): in our example, the stress-energy tensor is a functional $t:{\cal S}\rightarrow\mathbb{R}$, hence the notation $t[F]$. Finally, the dynamics, ${\cal D}$, is summarised by a set of two equations: namely, the 2-form $F$ is both closed and co-closed, i.e.~$\dd F=\dd*F=0$ (hence it is harmonic). The symmetries of this theory are Poincar\'e transformations, which map $F$ and $t$ to themselves (of course, their components transform covariantly).

This example prompts the following remark about theories thus defined: notice that the state-space of Maxwell's theory could equally well be presented in terms of the electric and magnetic fields, rather than the Faraday tensor. Thus there is no claim here that there is always a unique best definition of a theory, and choices are required about what to take as the state-space, and even judgment about interpretative matters. We will see another example of this in Section \ref{EMT}.





Recall that, as I announced at the start of Section \ref{thmod}, a model is in this paper not a particular solution. Rather, a {\it model} $M$ of a bare theory, $T$, is a realization, or mathematical instantiation: i.e.~it is a mathematical entity having the same structure as the theory, and usually some specific structure of its own. More specifically, I will use a notion of model as a representation (in the mathematical sense, not the philosophical one!) of the theory, i.e.~a homomorphism from the theory to some other known structure---think, for example, how an abstract group can be represented by a set of $n\times n$ matrices, where the rank, $n$, is specific structure that goes beyond the definition of (the homomorphic copy of) the group. 

But a model must not be a merely mathematical representation, for I need to make the following physical distinction within the homomorphism. A model of a physical theory naturally suggests the notions of:

A {\it model root}: the realization of the theory, usually its homomorphic copy (though, in some cases, an isomorphic copy).

The {\it specific structure}: that structure which goes into building the model root, which is not part of the theory's defined structure, and which gives the model its specificity. Specific structure may well be physically significant, depending on the context of application of the theory: and part of it is normally used for calculations within the model---but calculations can of course be done in different ways, using different specific structure. For examples, see: Sections \ref{gom}, De Haro (2016:~\S2.1), and De Haro and Butterfield (2017:~\S5.2).

It is helpful to have a schematic notation for models that exhibits how a model augments the homomorph of the structure of a bare theory, $T$, with its own specific structure: 
\bea\label{MmM}
M=\bra m;\bar M\ket~.
\eea
Here, $m$ is the model root, and $\bar M$ is the specific structure which goes into building $m$. In cases where $T$ is a triple, $m$ must itself be a triple with properties that are homomorphic to those of $T$. When the model root, $m$, is itself a triple, then we call it the {\it model triple}.\footnote{Notice that the distinction between the model root and the specific structure, even if formalised into the definition of the model as in Eq.~\eq{MmM}, is conceptual rather than strictly mathematical. This means, firstly, that it can only be made on a case-by-case basis; and second, that in Eq.~\eq{MmM}, we should not think of $m$ and $\bar M$ as being given independently and together defining $M$. Rather, $\bar M$ is the specific structure from which the model root $m$ is built.}

Like bare theories, models (and model triples) are, at this stage, {\it uninterpreted}; and an interpretation can again be added as a set of partial maps (see Section \ref{seminter}).

We will sometimes need a notation for the model as a homomorphism, from the theory to the structure $m$ that does the (mathematical!) representing. I will denote it as follows:
\bea\label{hmtm}
h_m:T\rightarrow m~,
\eea
where $m$ labels the representation. As usual, there is a triple of such maps, one for each item of the theory. For examples of models as representations in Maxwell's theory, see Section \ref{emm}.


\subsubsection{Interpretation}\label{seminter}

I will adopt a mainstream position about interpretation: namely, classical referential semantics. Thus we will assign references in the empirical world to the elements of a theory---viz.~to the states and the quantities, and to the formulas built from them---and likewise for models.\footnote{Referential semantics, at least in its mainstream presentation, may admittedly have some limitations in that it sometimes does not pay sufficient attention to scientific practices, norms, and non-linguistic skills (as witnessed by works, such as Montague (1970), that are supposedly about ``pragmatics'' in the philosophy of language, but remain largely formal). For an interesting critique of such an ``ideal of pristine interpretation'' (admittedly, somewhat caricaturized), see Ruetsche (2011, pp.~3-4). However, there is {\it no incompatibility} between using a referential semantics for theories, and other lines of work that emphasise complementary aspects such as practices, norms, and non-linguistic skills. For a defence of this irenic perspective, see Lewis (1975:~p.~35), De Haro and Butterfield (2018:~Section 2.1.3), and De Haro and De Regt (2018:~pp.~633-637).} One advantage of the framework of referential semantics is that both realists and constructive empiricists can agree about the interpretation of a theory or model, in other words about its basic ontology (`the picture of the world drawn by the theory', to use van Fraassen's (1980:~pp.~14, 43, 57) words), even though they have different degrees of belief in the entities that the ontology of the theory postulates---and indeed they will have a different conception of what such a map entails. We will see illustrations of this point below, and in Section \ref{comptheq}.

Thus I will model referential semantics by interpretation map(s), as follows:

An {\it interpretation} is a set of partial maps, preserving appropriate structure, from the theory to the world.\footnote{I will occasionally use the phrase `physical interpretation', in order to distinguish this interpretation (i.e.~what the items of the theory refer to in the world) from the linguistic interpretation (i.e.~the language of---an abstract form of---mathematical physics in which the theory is formulated). The condition that the map only needs to be partial means that there need not be a reference for all its arguments, i.e.~for all the terms in the bare theory. This is because some interpretations may map to worlds with less structure than envisaged by the bare theory. See De Haro and Butterfield (2018:~\S2.4). However, this point will not be very relevant in this paper.} The interpretation fixes the reference of the terms in the theory. More precisely, an intepretation maps the theory, $T$, to a domain of application, $D$, in the world, i.e.~it maps $i:T\rightarrow D$. If the theory is a triple of states, quantities, and dynamics, viz.~$T=\bra{\cal S},{\cal Q},{\cal D}\ket$, then the interpretation maps are also a triple, one for each item in the triple: however, we will not often need to make this explicit. Using different interpretation maps, the same theory can describe different domains of the world, and even different possible worlds. For more details, also about the kinds of maps required, see De Haro (2016:~\S1.1.2).

Likewise for models, an interpretation is a partial map, $i:M\rightarrow D$, preserving appropriate structure, from the model to the world.

Notice that the choice of `the part of the structure that is common to all of the theory's models' goes into the definition of a model, and singles out the core physics that is described by the theory, because it is represented in all its models: namely, it is in the distinction between the model root and the specific structure from the previous Section: and this choice constrains the kinds of systems the model is able to describe. The way to determine the relevant structure is interpretative not formal. For example, only experiments can tell us that massless vector fields correctly describe photons. Once that question is experimentally settled, one can take a model root including a massless vector field to describe a photon. But once we have that model root with its massless vector field, we can strip it of its photon interpretation and use the same model root to describe whichever other particle exhibits degrees of freedom of the same kind. 

Interpretations thus defined are very general. To specify them further, we endorsed, in De Haro and Butterfield (2018:~Section 2.3), De Haro (2016:~\S1.1.2) and De Haro and De Regt (2018:~p.~636), a more specific framework, viz.~intensional semantics (cf.~Lewis (1970), Carnap (1947:~pp.~177-182; 1963:~pp.~889-908)). In this framework, the notion of `linguistic meaning' (for us, in the context of scientific theories: an `interpretation') is taken to be ambiguous between what Frege (1892) called `sense' and what he called `reference', here called `intension' and `extension' respectively. The intension is the linguistic meaning of a term, while an extension is the worldly reference of the term, relative to a single possible world (with all of its contingent details). Thus Lewis (1970:~pp.~22-27) defines intensions as maps from $n$-tuples of sequences of items---he calls such a sequence an `index'---to extensions, e.g.~the truth-values of sentences. And, as Lewis remarks, the framework also applies if the indices are construed as models consisting of states representing possible worlds (ibid, p.~23): see also Carnap (1947:~pp.~177-182; 1963:~pp.~889-908). Here, I will discuss a variation of this framework that involves: (a) mapping from scientific theories and models (usually presented as triples, so that there are three such maps) rather than from sequences of general linguistic items; (b) a simplification: namely, modelling both intensions and extensions by interpretation maps, rather than defining an extension as the (set of) objects/truth value(s) and an intension as a map to extensions. 

Thus both intensions and extensions are structure-preserving partial maps from a bare theory or model to a domain of application relative to a possible world. The difference between the intension and the extension is in the kind of domain of application: explicitly, for states: an intension maps a state to a generic property (or physical arrangement) of a system mirroring the defining properties of the mapped state (and likewise for quantities and dynamics). Thus the image of the state and the domain of application abstract from contingencies such as the detailed arrangement of the system and how the system is measured (so that the interpretation applies to all possible worlds that are described by the theory or model). By contrast, in the case of an extension, the image and domain of application are a fully concrete physical system: usually including also a specific context of experiment or description, and all the contingent details that are involved in applying a scientific theory to a concrete system (for more details, see De Haro and Butterfield (2018:~\S2.3) and De Haro and De Regt (2018:~\S1.1)). \\
\\
Let me illustrate this notion of interpretation in the simple case of Maxwell's theory in vacuum, and on $\mathbb{R}^4$, mentioned in Section \ref{thmod}: where the state-space ${\cal S}$ is a suitable set of 2-forms $F$, the quantities ${\cal Q}$ contain a distinguished quantity, the stress-energy tensor, with components $T_{\m\n}[F]$, and the dynamics is the condition that the 2-forms $F$ are both closed and co-closed (more details in Section \ref{Maxem}). Under the standard electromagnetic interpretation, the Faraday tensor $F$ is the coordinate-free presentation of a tensor whose components are interpreted as electric and magnetic fields, so that the intensional interpretation map is:
\bea\label{i(F)}
i(F)&=&\mbox{`a coordinate-free specification of an electric and magnetic configuration}\nn
&&~~~~~~~~~~~~~~~~~~~~~~~~~~~~~~~~\mbox{in vacuum'}~.
\eea
The key interpretative parts are here `coordinate-free specification' and `electric and magnetic configurations', which we (uncontroversially) assume already have established meanings, for example as correlating with certain experimental procedures that we use to measure the fields, or as corresponding to certain properties of fields that we are familiar with in our world (such as the polarisation of light waves, and how they interact with other entities). 

This meaning has come about through centuries of experimentation, instrumentation, and theorising about electro-magnetic fields. Different historical epochs may of course have different ways of manipulating and measuring these fields, but referential semantics assumes such reference to be clear, the more because the theory is sufficiently well established, as Maxwell's electromagnetic theory indeed is---and this reference is construed as saying that there are such things as electric and magnetic fields in the world (if one is a realist) or that there are indeed appearances of phenomena of electric and magnetic fields (if one is an empiricist). Likewise for the phrase `coordinate-free specification': it summarises the properties of the fields under changes of the frame of reference, i.e.~properties like `the electric field Lorentz is augmented in the directions perpendicular to the motion by an amount given by the Lorentz factor'. Such properties again uncontroversially correlate with particular phenomena in the world.\footnote{Interpreting $F$ in a coordinate-free manner emphasises the map's respecting the symmetries, which is one of the defining properties of interpretation maps. But the various components of $F$ of course also have their own interpretations, once a system of coordinates has been chosen. Thus $F_{i0}$ (where $i=x,y,z$) is interpreted as the electric field along the $i$-direction, and $F_{ij}$ is interpreted as the magnetic field perpendicular to the $ij$-plane.}

Notice the following two properties of the interpretation map Eq.~\eq{i(F)}. First, recall that I defined interpretation maps $i:M\rightarrow D$ as appropriately structure-preserving. Therefore the phrase `coordinate-free specification' in Eq.~\eq{i(F)} implements this appropriately structure-preserving character of the Poincar\'e symmetries. Namely, Poincar\'e symmetries, which leave the 2-tensor $F$ invariant and transform its components according to the standard Poincar\'e transformations, have a ``shadow'' in the domain of the world, $D$: this shadow contains the ordinary effects, of Lorentz expansions and contractions of the fields, that I referred to in the previous paragraph. Thus the interpretation map Eq.~\eq{i(F)} correctly preserves symmetries. Second, the interpretation Eq.~\eq{i(F)} is an intension: for it is valid at any possible world that instantiates electric and magnetic fields whose corresponding Faraday tensor satisfies the definitions of our states (viz.~a square-integrable, smooth 2-form). In other words, it does not depend on a specification of $F$---as being, for example, a collection of travelling waves with certain polarizations. Such ``generic'' interpretations hold at every possible world that is described by the theory. In order to get an extension, we should specify the particular 2-form $F$ that we are mapping. This would entail giving the details of its functional form: which then corresponds, under the interpretation map, to further specifications in a particular world, like `a wave travelling in such and such direction, with polarizations at such and such angles'. Since intensions are less detailed and simpler to write down in words, in the rest of the paper our examples will be mainly intensions. But, as I mentioned above, this way of modelling intensional semantics accommodates for boths kinds of interpretations.

There is also an interpretation map for quantities, which for example maps the 00-component of the stress-energy tensor:
\bea\label{emE}
i(T_{00})&=&\mbox{`the electromagnetic energy density of the system'}~.
\eea
Again, this map is an intension---it holds at all possible worlds at which Maxwell's theory applies, and in which a frame of reference has been specified, relative to which the 00-components of the tensor are taken. To specify the extension of $T_{00}$, we should provide additional details such as where the electromagnetic energy is produced from, how it interacts with its environment and is measured, etc. We will return to the example of Maxwell's electromagnetic theory in Section \ref{Maxem}.\\


So much by way of introducing referential semantics and intensional semantics. The idea of theoretical equivalence will be, roughly, that two models are isomorphic relative to their formal structure---as in their model roots, $m$---and their interpretation. Thus I will now define an interpretation that depends only on the model root, and not the specific structure:\\
\\
{\bf Internal interpretation:} an interpretation that maps all of and only the model root, regardless of the specific structure of the model. 
Since internal interpretations map the model roots (and they may map specific structure only in so far as it appears in the model root), it will often be clearer to restrict the internal interpretation map of models to the model root, and write: $i:m\rightarrow D$. Internal interpretations obviously also apply to bare theories, $T$, which do not have any specific structure.

I shall contrast this with other interpretations that also map the specific structure, and which I dub the {\it external interpretations} (for a fuller exposition, see De Haro (2016:~\S1.1.2)). For the reason given above, in the rest of this paper the relevant interpretations will be only the internal interpretations.\footnote{The reason to consider only internal interpretations in analysing theoretical equivalence can be spelled out as follows. Any two isomorphic but distinct models have different specific structure. Thus any pair of external interpretations that interpret the specific structures of the two models differently would render the models theoretically {\it inequivalent}. If one is interested in theoretical {\it equivalence}, it therefore makes good sense to restrict the interpretations to those that interpret only the isomorphic structure---which also rids one of the potential arbitrariness of counting as theoretically inequivalent those models that differ only by a ``trivial convention'', if their specific structure differs. The distinction between the model triple and the specific structure aims to make a distinction between, roughly speaking, what is the physical core of the model and what is additional structure (which may be physical in some contexts but not in others). See also the wise remarks, in Butterfield (2018:~\S3.3), about the flexibility of the distinction between `fact' and `convention'. This is, of course, not to say that two models could not be theoretically equivalent on external interpretations: but only that we get a natural meshing condition between theoretical equivalence and duality, as we should, {\it if} we consider internal interpretations: as I will show in Section \ref{sdc}.\label{whyint}} 

I use `internal interpretation' in the singular here because we will normally consider a single interpretation. However, a given model can have multiple internal interpretations. For example, take a model whose dynamics is the heat equation for a real, non-negative function over space and time that we denote by: $\r:\mathbb{R}^4\rightarrow\mathbb{R}_{\geq0}$, and is given by: $(t,{\bf x})\mapsto\r\,(t,{\bf x})$. Take the state-space of the model to be the space of configurations of the function $\r(t,{\bf x})$ (and let the quantities of the model also be constructed from $\r$, its powers, and its derivatives and integrals over space and time---so that we now have a triple of states, quantities, and dynamics). This model can be interpreted as describing the spread of heat in time in a given region of space, or as a diffusion equation for Brownian particles in a medium. With these definitions, there is here a single model (in fact, a theory), with no specific structure. Both interpretations are internal, and yet they differ.\footnote{The current formulation of the internal interpretation, following De Haro (2016:~\S1.1.2), maps only the model root: i.e.~the interpretation map, which is a partial map, is {\it not defined} on the parts of the specific structure which are not part of the model root. This gives a clear-cut criterion for when an interpretation is internal. The formulation in Dieks et al.~(2015:~pp.~208-210) might, on the other hand, give the (mistaken) impression that an `internal viewpoint' somehow requires an interpretation that is internal {\it to the theory} and is constructed with no other information. But this is of course not how interpretations come about in physics (as I also emphasised in the example of Maxwell's theory): they are always constructed against the background of already existing theories. This methodological point is expounded in De Haro and De Regt (2018), which gives three interpretative tools that can be used to construct {\it internal} interpretations, i.e.~without interpreting the specific structure (see also Dewar (2017:~\S6) for some excellent comments on related issues). One of the three tools is indeed internal to the {\it theory}, but two others connect to already existing theories, while still being `internal' in my sense. Thus, just as dualities are used to internally interpret theories, other inter-theoretic relations are also used. In conclusion, the meaning of `internal' is not `internal to the model' but `internal to the model root' (or the common core, in cases of duality).}

I also assume that the model root is rich enough that an internal interpretation allows the specification of a set of possible worlds that instantiate the solutions of the equations of the model (or, at least, the interpretation specifies suitable domains of application within such worlds: see the discussion in Section \ref{comothers}, especially footnote \ref{RMN}).  
Nevertheless, for simplicity, I will talk about `the possible world specified by the model' rather than about the set of such worlds.



\subsubsection{Duality and physical equivalence}\label{0dual}

In this Section, I give brief construals of the notions of duality and of physical equivalence according to our proposed Schema for duality (and I shall elaborate on physical equivalence in Section \ref{diffeq} after I define, in Section \ref{nteq}, weak theoretical equivalence). 

Indeed, having discussed theories and models, we can now easily state the Schema's conception of duality: {\it A duality is an isomorphism between model roots of a single bare theory, where the model roots are taken to be representations (in the mathematical---not semantic!---sense) of the theory.} We denote this isomorphism of model roots, $m_1\cong m_2$, by the map $d:m_1\rightarrow m_2$.

Theories may have many representations: representations that are isomorphic to the original theory and representations that are not isomorphic to the original theory. But if we have two (or more) representations that are isomorphic to each other (whether they are isomorphic to the original theory or not), we have a duality. \\

\begin{figure}
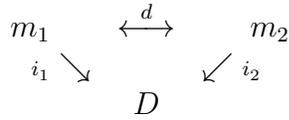

\begin{center}
\bea
\begin{array}{ccc}m_1\!\!&\overset{d}{\longleftrightarrow}&\!m_2\\
~~~~~~~{\sm{$i$}}_{\tn{1}}\searrow&&\swarrow \sm{$i$}_{\tn{2}}~~~~~~~~\\
&D&\end{array}\nonumber
\eea
\caption{Physical equivalence. The two interpretations describe ``the same sector of reality'', so that the ranges of the interpretations coincide.}
\label{dual3}
\end{center}
\end{figure}

The framework offered so far is mostly formal. Thus it is not surprising that the Schema contributes to furthering understanding of formal equivalence---what I will call {\it weak theoretical equivalence}. This will be the topic of Section \ref{gdte}. But first, by way of introduction, let me say a few words about {\it physical equivalence,} i.e.~the specific proposal for theoretical equivalence suggested by recent developments in physics---a fuller exposition is in Section \ref{diffeq}.\footnote{The notion of {\it physical equivalence} that I use here, and that I discuss more fully in Section \ref{diffeq}, has been explicated in more detail De Haro (2016:~\S1.3).} Physical equivalence can be defined as sameness of interpretations, i.e.~two models are physically equivalent iff they are weakly theoretical equivalent and, in addition, they {\it have the same interpretation} (not in the sense of having the same interpretation maps, but in the sense that their interpretation maps have the same codomain, or target, and appropriately matching values in that codomain). 
Thus physical equivalence, for dual models, is a meshing condition between duality and intepretation: namely, in the notation of Figure \ref{dual3}, $\mbox{ran}\,(i_1)=\mbox{ran}\left(i_2\circ d\right)$. Duality and interpretation then commute, in the sense that they form a triangular commuting diagram.

\subsubsection{Against na\"ive appeals to isomorphism}\label{agiso}

In this Section, I elaborate on the notion of isomorphism employed by the Schema, and explain how it differs from na\"ive notions of isomorphism.

Whether a map is an isomorphism of course depends on the structure that the map is supposed to preserve. In the present case, of isomorphism between model roots, it is the structure of the model roots---and so of the bare theory of which each root is a representation---that is preserved. And when the model roots are triples of state-space, quantities and dynamics, then each of these three items---with their corresponding structure---is to be preserved.

This contrasts with a vague, and-or unqualified, notion of isomorphism. The isomorphism criterion of models is usually associated with the semantic view: see, for example, Suppe (2001:~p.~526) and Halvorson (2012:~p.~187). For example, Halvorson (2012:~p.~188) has summarily criticised the isomorphism account of theoretical equivalence on the grounds that it would render Heisenberg's matrix mechanics and Schr\"odinger's wave mechanics theoretically inequivalent, for: `a matrix algebra is obviously not isomorphic to a space of wave functions; hence, a simpleminded isomorphism criterion would entail that these theories are inequivalent.' 

The first point to make is that Halvorson slips in his formulation of the claim he is targeting. For of course the matrix algebra represents the quantities, while the wave-functions form a space of states---and in that sense I agree that they should not be isomorphic. But, correcting this slip: I of course agree with Halvorson that the representation space of a matrix algebra is in general not isomorphic to a space of wave-functions, and so the isomorphism criterion does not work in this case. But this is only a na\"ive notion of isomorphism, which attempts to map the state-spaces directly onto each other, so that discrete matrix indices are directly mapped to the state-space of continuous functions (von Neumann (1955:~p.~22)). Such an isomorphism requires the introduction of what von Neumann called Dirac's `mathematical ``fiction''', namely the Dirac delta distribution: `this cannot be achieved without some violence to the formalism and to mathematics' (ibid, p.~28).\footnote{For a philosophical and historical exposition, see Muller (1997).} Setting aside the question whether Dirac's programme can perhaps be carried out using the theory of distributions: one mathematically correct procedure is to seek an isomorphism, not of configuration spaces, but of {\it functions on these two spaces}, i.e.~between sequences  which are functions on the discrete configuration space, $l^2(\infty)$, and wave functions on the continuous configuration space. It is {\it this} weaker isomorphism that quantum mechanics requires: `These functions... are the entities which enter most essentially into the problems of quantum mechanics' (ibid, p.~28). And those two spaces {\it are} isomorphic, as a theorem of Fischer and Riesz secures (cf.~e.g.~von Neumann (1955:~p.~29)). Indeed, all complex infinite-dimensional separable Hilbert spaces are isomorphic to $l^2(\infty)$ (e.g.~Prugove\v{c}ki (1981:~p.~41)). 

It seems likely that the claims, now well-rehearsed, in the literature (see North (2009: p.~84-88), Curiel (2014:~pp.~295-304), Barrett (2018:~\S1)), that Lagrangian and Hamiltonian mechanics are inequivalent according to the isomorphism criterion, can be dealt with in a similar way, i.e.~that a more sophisticated notion of isomorphism does render them equivalent, at least in an important class of salient cases. The leading idea of any such isomorphism must of course be the Legendre transformation, which, as textbooks teach us, carries one back and forth between the Lagrangian and Hamiltonian frameworks. For example, Barrett (2018:~\S4) focusses on the case where the Legendre transform is a diffeomorphism (the Lagrangian then being called `hyperregular'). Analogously to the case of matrix vs.~wave mechanics, the criterion would obviously {\it not} be a direct isomorphism of tangent and cotangent bundles, which are {\it not} invariably isomorphic.

Similarly, Lutz (2017:~pp.~335-336) proposes to block a charge against the na\"ive isomorphism criterion\footnote{This time, the charge (Halvorson (2012:~p.~190)) is rather that the isomorphism criterion fails to distinguish different theories.} by considering a more sophisticated criterion of isomorphism: namely, he considers indexed structures and their associated isomorphisms. 

Without going into further details here: the target of the criticisms of the isomorphism criterion for theoretical equivalence often seems to be a na\"ive, i.e.~vague and-or unqualified, appeal to isomorphism, which either does not specify the structure to be preserved, or identifies the wrong structure. On a more sophisticated treatment of the structures involved---which often means: a more restrictive definition of the relevant structure to be preserved---cases of theoretical equivalence in physics {\it do} often seem to be cases of isomorphism.

Therefore, I find most of the reasons in the recent philosophical literature, for abandoning the isomorphism criterion, to be wanting. 
The isomorphism criterion can be upheld provided we are more careful about the structures that we claim are isomorphic. It is in this spirit that I will propose an isomorphism criterion in the next Section.

\subsection{The Schema's contribution to a conception of theoretical equivalence}\label{nteq}

This Section and the next contain the paper's main proposal for theoretical equivalence, based on the Schema. In Section \ref{gdte}, I give some background on the notion of theoretical equivalence in recent philosophy of science discussions, in terms of two conditions. Section \ref{scte} then goes on to fill in the first condition according to the Schema, and Section \ref{sdc} fills in the second condition. 

\subsubsection{The general notion of theoretical equivalence}\label{gdte}

In this Section, I discuss the general idea of theoretical equivalence in recent philosophy of science. Theoretical equivalence is, in fact, something of a term of art, without a fixed meaning. It usually combines two conditions:\\
\\
(A)~~some sort of ``formal or mathematical requirement''; and \\
(B)~~some sort of ``interpretative requirement''.\\
\\
Individual authors may of course stress one aspect over the other, or even altogether reject one of the two conditions.\footnote{For example, Coffey (2014:~p.~837) proposes an account in which questions of theoretical equivalence ultimately reduce to questions of interpretation, and formal considerations are only relevant in virtue of how they shape interpretative judgments.}

Two influential proposals for theoretical equivalence are due to Quine and Glymour. On Quine's (1975:~p.~320) proposal, two formulations count as formulations of the same theory if, besides being empirically equivalent, the two formulations can be rendered identical by a {\it reconstrual of the predicates} in one of them (for example, by switching some of the predicates). And Glymour (1970:~p.~279) requires, besides empirical equivalence, {\it inter-translatability} as a necessary condition for theoretical equivalence.\footnote{For how Quine's and Glymour's views on equivalence relate to the question of empirical under-determination, see De Haro (2019:~Section 2.3).} Thus Quine and Glymour agree that empirical equivalence is a necessary condition for theoretical equivalence, but construe the formal requirement (reconstrual of predicates vs.~inter-translatability) differently. And both do seem to have in mind a notion of theoretical equivalence in which both theories in some sense ``say the same thing about the world'', i.e.~a notion that is also interpretative.\footnote{However, one should note that Quine's and Glymour's interpretative attitudes are quite different, because Quine was sceptical about meaning and ontology (see Quine's (1960:~Section 2) discussion of referential indeterminacy) while Glymour (1977:~p.~228) is not so sceptical.}

About `empirical equivalence': there are two main construals of the requirement of this criterion, one of which is syntactic, and the other semantic. Quine (1970:~p.~179, 1975:~p.~319) says\footnote{Glymour (1970~:~p.~277) holds a similar view.} that two theories are empirically equivalent if they imply the {\it same observational sentences,} also called `observational conditionals', for all possible observations---present, past, and future. Van Fraassen (1980:~p.~64) says that two theories, $T$ and $T'$ are empirically equivalent if for every model (not the Schema's model!), $M$, of $T$ there is a model, $M'$, of $T'$ such that all of $M$'s {\it empirical substructures are isomorphic} to empirical substructures of $M'$, and vice-versa.\footnote{For a comparison between Quine's syntactic and van Fraassen's semantic conceptions of empirical equivalence, see De Haro (2019:~Section 4.2).} The criterion of theoretical equivalence that I will give in Section \ref{diffeq} entails empirical equivalence as a special case (for a detailed discussion, see De Haro (2019)). 


The recent discussion of theoretical equivalence by Halvorson (2012), Barrett and Halvorson (2015, 2016), Halvorson and Tsementzis (2015), Barrett (2018), and Weatherall (2015, 2016, 2016a) has emphasised the formal aspects of equivalence over the interpretative. 
Weatherall mentions empirical equivalence in all of his papers on the topic, but for example Barrett and Halvorson (2015, 2016) do not mention it at all\footnote{Barrett (2018) and Rosenstock et al.~(2015) also mention empirical equivalence.}---nor do those papers mention matters of meaning or interpretation (other than interpretation using formulas in a formal language). Thus one could easily get the impression that some of these authors ``are not interested in interpretation''. However, the situation is more subtle.\footnote{In conversation, Weatherall and Barrett say both that empirical equivalence is a prerequisite for their analyses of theoretical equivalence, and that a suitable notion of theoretical equivalence should (at least) respect (some) meanings, i.e.~it is constrained by the interpretation. However, the latter requirement is not mentioned in their papers, and it is not clear what guiding role it plays. This contrasts with the older literature on the topic. For Glymour's (2013:~p.~289) and van Fraassen's (2014:~pp.~278-281) comments about interpretation already go some way towards the kind of interpretative project I have in mind here. Their comments of course go back to their own accounts of theories and of theoretical equivalence, in Glymour (1970, 1977) and van Fraassen (1970).}

I will endorse this recent (though usually unstated) consensus, that there is an interesting part of the project of theoretical equivalence that is a {\it largely, but not solely, formal matter.}\footnote{Butterfield's (2018:~Section 1.1) {\it Remark}, namely, `in physics, two theories can be dual, and accordingly get called `the same theory', though we interpret them as disagreeing', emphasises the distinction between the formal and the interpretative aspects. In the context of theoretical equivalence, Butterfield's {\it Remark} has also been voiced by van Fraassen (2014:~pp.~278-279); and, in the context of duality, by De Haro (2016:~\S1.3-\S1.4, \S2.2).} Thus I will first develop, in Section \ref{scte},  a notion of `weak theoretical equivalence' (I will discuss physical equivalence in Section \ref{diffeq}). Indeed I believe (contra Coffey (2014)) that there is an interesting philosophical project of:

(EE)~~{\it Explicating equivalence in formal terms}.

This consensus position in the recent discussion of theoretical equivalence can be qualified as a {\it quietist} position. Namely, it is the position of authors engaged with parts of the project of theoretical equivalence, and for whom interpretative equivalence is a minimal requirement that their project needs, but on which they do not wish to focus. And so, such authors, when confronted with dual models, discuss formal equivalence but typically refrain from discussing ontological questions. Quietism is the position that my account of weak theoretical equivalence attempts to save, by including an interpretative requirement that is as minimal as possible---namely, mere matching of interpretations. The account can be suitably strengthened in the second step, as I will do in Section \ref{diffeq}. Such authors have nothing to say about physical equivalence---for whatever reason: perhaps because they are sceptical about the notion, or not interested in it, or think it is too difficult to articulate. This is, to a large extent, a tenable position:\footnote{The reason for the qualification `to a large extent' is that it is not automatic that the various extant formal proposals---definitional equivalence, Quine equivalence, Morita equivalence, categorical equivalence---will give physically equivalent theories. However, in Section \ref{diffeq} I will argue that, given weak theoretical equivalence, one can construct interpretations that deliver physical equivalence.} and in a moment I will add some of my own reasons why (EE) is an interesting philosophical project.

In light of the literature discussed above, the challenge in the first step of defining weak theoretical equivalence lies, for the most part, in how one fills in its being `a largely, but not solely, formal project'. That is, the difficulty lies in striking an appropriate balance between the interpretative and the formal: in such a way that, once one takes the interpretative conditions to have been established for a pair of theories, as a kind of ``boundary condition'', then the project can concentrate on the formal issues.

In the next Section I will propose how the Schema suggests this can be done. But it will clarify that proposal if I first give three reasons why I think `weak theoretical equivalence', in contrast with `physical equivalence', should be construed as being {\it mostly} formal or mathematical: \\
\\
(i)~~~The notion of weak theoretical equivalence will easily lead in to a notion of physical equivalence, which can then be further analysed.\\
\\
(ii)~~The second reason for letting weak theoretical equivalence be formal and not require sameness of interpretation is that establishing sameness of interpretation is never a simple matter: for it requires a notion of identity of domains of application, as I will discuss in Section \ref{diffeq}, which leads in to issues of metaphysics. And we do not want to be too worried about metaphysics in our project of theoretical equivalence. 

I take this second reason to underlie the recent consensus mentioned above: that there is a significant conceptual and technical project of finding criteria for when two theories are equivalent, without the need for all the details about the interpretation to be fleshed out, nor to commit to a metaphysical account of how the terms of the theory refer. 
And such a project strikes me as sensible.

This then prompts me to break up the overall project of explicating equivalence between theories into two tasks. The first task, namely explicating {\it weak theoretical equivalence}, cares as little as possible about interpretation: and it corresponds to (EE) above. The second task, namely explicating {\it physical equivalence}, takes interpretation fully on board.\\
\\
(iii)~~~~The third reason for keeping weak theoretical equivalence largely formal is that, as the Schema for duality has shown, this is indeed also how physicists think about equivalence (cf.~De Haro (2018:~p.~\S4.1.2)). Physicists are happy to say that the high-temperature Ising model is equivalent to a low-temperature one, without worrying about the ontological status of `temperature' under such a duality map. It suffices that two variables that are interpreted as `temperatures' are mapped to each other, and that the values (low vs.~high) of `temperature' are also mapped to each other. Even less do physicists worry whether they are e.g.~realists or empiricists about temperature under this duality. And this is of course a legitimate practice. Indeed I of course maintain that the metaphysical questions are important; but they can be addressed in the next step, as part of the project of {\it physical} equivalence. Thus a project like (EE), that only takes interpretation minimally into account, agrees with a widespread, and justified, scientific practice.\\

To sum up: regardless of one's preferred use of `theoretical equivalence': it is an interesting project to establish criteria of formal equivalence that minimize interpretative issues, and so avoid being committed to the fully-fledged project of showing that the theories have the same interpretation (for they may {\it not} have the same interpretation!): and this is what I shall call the project of weak theoretical equivalence.

But if one endorses this recent consensus---that weakly theoretically equivalent models are {\it in some (weak!) sense} interpretatively equivalent: but that establishing weak theoretical equivalence does not require one to have a complete understanding of the interpretation, so that the project is largely formal---then it is clear that achieving what the consensus aims for requires a delicate balance. 

So, it is now crucial to fill in condition (B)---the interpretative requirement---in such a way that we can strike this balance. I will propose that the models must have {\it matching internal interpretations}. 

\subsubsection{Condition (A'): isomorphism of model roots}\label{scte}


I first fill in condition (A) in Section \ref{0dual}: namely, the formal or mathematical requirement for weak theoretical equivalence, which I will illustrate in Section \ref{teMx}.
My proposal here is simple: namely, to read the Schema's notion of duality as a {\it formal condition of theoretical equivalence,} applicable to formal theories in the physical sciences:\footnote{Notice that this proposal can be discussed independently of whether one endorses my explication, (B'), of (B): the proposal could be combined with any other explication, (B'').}\\
\\
(A')~~{\bf Isomorphism of model roots of a single bare theory.} \\
\\
Recall that this was precisely the definition of duality, in Section \ref{0dual}. And it is indeed natural to regard duality as prompting the weak theoretical equivalence of two models (in my sense of `model'). 

\subsubsection{Condition (B'): matching internal interpretations}\label{sdc}

In this Section, I fill out the second condition for weak theoretical equivalence: namely, the interpretative requirement, as prompted by the Schema's analysis of duality. 

We have seen that the spirit of weak theoretical equivalence is that it should capture when two interpreted models (in the sense of model roots, i.e.~often called `theories') 
are equivalent, both formally and interpretatively, without being committed to a strong notion of sameness of interpretation.
This is the `delicate balance' that I mentioned at the end Section \ref{gdte}.  

My proposal, then, for filling in condition (B), namely filling in the ``interpretative'' part of the definition, is to require ``matching interpretations''. So, let me first define this notion: this will lead us to the restriction to internal interpretations (as defined in Section \ref{seminter}).\\
\\
(B')~{\bf Matching interpretations:} two models, $M_1$ and $M_2$, have matching interpretations when the ranges of their interpretation maps are isomorphic, i.e.~when:
\bea\label{equivI}
\mbox{ran}\,(i_1)\cong\mbox{ran}\,(i_2)~.
\eea
I use the word `matching' here, in order to avoid the already over-used word `equivalence'. Recall, from Section \ref{seminter}, that an interpretation is a structure-preserving partial map that maps a model to a domain of application. For the two models $M_1$ and $M_2$, we thus have maps $i_1:M_1\rightarrow D_1$ and $i_2:M_2\rightarrow D_2$. Here, the domains of application $D_1$ and $D_2$ are structured sets, and the interpretation maps are structure-preserving. Thus, the condition Eq.~\eq{equivI} is an isomorphism with respect to that structure, which is induced from the model's own structure. 


For simplicity in the discussion in the rest of this Section, I will take the interpretations to be surjective maps, so that $\mbox{ran}\,(i_1)=D_1$ and $\mbox{ran}\,(i_2)=D_2$, i.e.~the interpretation maps map ``to the whole domain of application'': every element in the domain of application is mapped to by at least some element of the model. (This restriction does not affect the content of the discussion but does simplify the notation.) Thus, we can now restate the condition Eq.~\eq{equivI} as follows: $D_1\cong D_2$. Let us denote the corresponding isomorphism between domains of application by $\ti d$, so that:
\bea\label{tid}
\ti d:~D_1\rightarrow D_2~.
\eea
I shall call this map, induced from the duality and  interpretation maps, the {\it induced duality map}.

Combining the duality map, $d$ (defined in Section \ref{0dual}), with the two interpretation maps, we get the diagram in Figure \ref{eqmapI}, where the interpretation maps, $i_1$ and $i_2$, are evaluated on the model roots, $m_1$ and $m_2$ (see the discussion two paragraphs below). Here, $d$ and $\ti d$ are both isomorphisms, while the interpretation maps of course need not be. Figure \ref{eqmapI} thus contrasts with the diagram in Figure \ref{dual3}, where the duality {\it commutes} with the interpretation---in the sense that the three maps $d,i_1,i_2$ form a commuting diagram. In Figure \ref{eqmapI}, the three maps $d,i_1,i_2$ do not form a commuting diagram. Rather, we need to co-vary the duality map, thus getting the induced duality map, $\ti d$, in Eq.~\eq{tid}, as we change the interpretation.
\begin{figure}
\begin{center}
\bea
\begin{array}{ccc}m_1&\overset{d}{\longleftrightarrow}&m_2\\
~~\Big\downarrow {\sm{$i_1$}}&&~~\Big\downarrow {\sm{$i_2$}}\\
D_1&\overset{\ti d}{\longleftrightarrow}&D_2
\end{array}\nonumber
\eea
\caption{Weak theoretical equivalence and the induced duality map, $\ti d$, between domains of application.}
\label{eqmapI}
\end{center}
\end{figure}

The notion of co-variation of the interpretations with the duality map is natural for dualities.\footnote{Read and  M\o ller-Nielsen actually define {\it theoretical equivalence} in this way, i.e.~as empirical equivalence plus duality.} For example, take Kramers-Wannier duality, i.e.~the Ising model with high, respectively low, temperature as its two models (cf.~Butterfield (2018:~\S4.3)). These two models are isomorphic, and furthermore we can also match their interpretations by everywhere replacing a lattice at high temperature with a lattice at low temperature. 
Thus the isomorphism, $d$, between the models induces an isomorphism, $\ti d$, between the domains of application, that takes `high temperature' to `low temperature' and vice versa, while respecting the syntax. The induced duality map between the domains of application, thus construed, is analogous to translating one language into another. Of course, such a map does not in general {\it preserve} meanings---rather, it {\it maps} them into each other in a non-trivial manner!

Weak theoretical equivalence, as discussed here, is in the context of duality, i.e.~of an isomorphism account of equivalence. Thus in particular, the diagram in Figure \ref{eqmapI} implies that $i_2\,\circ\, d=\ti d\,\circ \,i_1$. Now, in order for $i_2$ to be defined on, or ``match'', the range of $d$ (i.e.~the model root, since the duality map only maps the model roots, see Section \ref{0dual}), $i_2$ must be restricted to the model root, and likewise for $i_1$ (as shown in Figure \ref{eqmapI}). This suggests that, for the purpose of theoretical equivalence between dual theories, we should restrict our interpretations to the {\it internal interpretations,} as I anticipated in Section \ref{seminter} (see especially footnote \ref{whyint}). For such a commuting diagram does not exist for external interpretations, as I now show.

In principle, weak theoretical equivalence could also be defined for external interpretations. We would then replace the model roots $m_1$ and $m_2$ in Figure \ref{eqmapI} by the full models, $M_1$ and $M_2$ (cf.~Eq.~\eq{MmM}), and $i_1$ and $i_2$ would be their external interpretations. We would then still require the map, $\ti d$, between the domains of application to be an isomorphism, as in Eq.~\eq{equivI}. But notice that the diagram is no longer commuting: for the cardinalities of $M_1$ and $M_2$ may well differ, and so $M_1$ and $M_2$ are in general {\it not} isomorphic.\footnote{If the full models, $M_1$ and $M_2$, including their specific structure, do happen to be isomorphic, then one could define weak theoretical equivalence for external interpretations as an isomorphism of the full models, i.e.~$M_1\cong M_2$. But then we are changing the notion of duality, which was tied to the notion of the common core theory, and we lose the connection between the common core theory and the duality. In such a case, we might say that the full models, $M_1\cong M_2$, {\it are} the common core, and that consequently there is no specific structure---but then we are back to the original case in the main text, of two isomorphic model roots. And then the external interpretations simply collapse to internal interpretations. And so, the only gain in defining duality as an isomorphism between full models $M_1$ and $M_2$ (in cases where such an isomorphism obtains), while still maintaining that the model roots are smaller than $M_1$ and $M_2$, is that the common core comes out to be smaller---thus, as far as I can see, no real gain. And so, the analysis in the main text is enough.} Thus, although weak theoretical equivalence {\it could} be defined for external interpretations as the isomorphism condition Eq.~\eq{equivI}, there is no commuting diagram criterion for it, so that this sort of weak theoretical equivalence is unrelated to the existence of duality. Thus in this paper we hold on to the connection between weak theoretical equivalence and duality, by restricting to internal interpretations (for the example of Maxwell's theory, see Section \ref{teMx}).\\

To sum up: I believe that making (B) precise as (B'), i.e.~matching of internal interpretations gives a minimal requirement of interpretative equivalence---as required for the `delicate balance' of the previous Section---that is useful both for dualities and for discussions of theoretical equivalence. In short: meanings are simply co-varied as we map one model to the other, without the requirement that the meanings stay ``the same''---the latter requirement is only introduced in the next step. Since the recent literature on theoretical equivalence has, as I mentioned, not given a conception of (B), I have here endeavoured to give a minimal one, based on the Schema. In the next Section, I will refine this to a criterion of theoretical equivalence.



Notice the difference between duality and weak theoretical equivalence thus construed: namely, condition (B') is required for weak theoretical equivalence, but not for duality. We follow the physicists in defining duality as isomorphism of model roots---period. Weak theoretical equivalence, on the other hand, is constrained by the philosophical tradition to require, in addition, an interpretative condition that I have here interpreted weakly, in the sense of Eq.~\eq{equivI} and Figure \ref{eqmapI}. Thus weak theoretical equivalence is an equivalence---an isomorphism---of interpreted model roots, while duality is an isomorphism of model roots, irrespective of their interpretation.

\subsection{Theoretical equivalence of duals, or physical equivalence}\label{diffeq}

In this Section, I spell out the notion of theoretical equivalence, or---in the context of physics, especially as suggested by dualities---{\it physical} equivalence: and discuss the relation between weak theoretical equivalence and physical equivalence. In Section \ref{comptheq}, I give sufficient conditions for physical equivalence, in the context of an isomorphism account of duality. In Section \ref{justify}, I discuss when a judgment of physical equivalence is justified, and draw lessons for theoretical equivalence in physics more generally. Then, in Section \ref{comothers}, I compare my account with other work.

\subsubsection{Physical equivalence}\label{comptheq}

Recall that I have explicated the weak theoretical equivalence of models as the matching of their interpretations. I will now define physical equivalence as weak theoretical equivalence plus, in addition, {\it sameness} of interpretation: more precisely, as sameness of the domains of application, $D_1$ and $D_2$ (cf.~Section \ref{0dual}, especially Figure \ref{dual3}). Thus the induced duality map, $\ti d$, between the domains of application, is the identity map: so that we have the commuting diagram, in Figure \ref{eqmapI}, between the duality and the  two interpretation maps.

But recall that, after making explicit the unstated consensus of the project of theoretical equivalence in Section \ref{gdte}, I noticed that this consensus aimed to stay formal and to minimize matters of metaphysics. And I justified this consensus as sensible, by using a notion of interpretation that assigned a basic ontology to the models, but remained quiet about other metaphysical matters. And that quietism is, in its turn, justified by intensional semantics' appropriately modelling the interpretative practices of both realists and empiricists: 
as in the interpretation maps that I defined in Section \ref{seminter}. Namely, recall the discussion in that Section: realists and constructive empiricists can agree about the interpretation of a theory or model, in other words about its basic ontology (`the picture of the world drawn by the theory', to use van Fraassen's (1980:~pp.~14, 43, 57) words), even though they have different degrees of belief in the entities that the ontology of the theory postulates. Also, realists may disagree amongst themselves about a metaphysical construal of those entities: think, for example, of Quine's (1960:~\S12) referential indeterminacy, according to which the linguist, upon hearing the native utter the word `gavagai' while pointing at a rabbit, might for simplicity translate it as `rabbit'---while still being at a loss whether the objects to which this term applies are rabbits, or stages, i.e.~brief temporal parts of rabbits, or mereological fusions of spatial parts of rabbits. 

In other words, whatever the formally coinciding domains of application are, one has to impose on both sides the same ontological construal of terms, e.g.~`energy is a property and not a relation', etc.

Thus, if {\it numerical identity} of the elements and relations in the domains of application---or, in a weaker sense, the {\it lack of a difference} between them---is what we want, we need an agreed philosophical conception of the ontology, i.e.~of the domains of application: which means that we now need to make our metaphysical commitments explicit, e.g.~about realism or empiricism and referential indeterminacy. This then amounts to a deeper explication of what we mean by an `interpretation' than just saying that it is a map. Only then does the requirement, that the induced duality map, $\ti d$ (Eq.~\eq{tid}), between the domains of application be the identity, secure this lack of a difference.

Thus I will say that the domains of application are (numerically) {\it the same} (in a weaker sense, that they {\it lack a difference}) if we have an {\it agreed philosophical conception of the interpretation}.\footnote{This is the content of the following remark in De Haro (2016:~Section 1.3.2): `Doing so [establishing physical equivalence] would require a deeper analysis of the notion of reference itself. Also, it is partly a metaphysical question whether, or under what conditions, for example two objects that are physical duplicates of each other (according to a given theory that describes all their properties) are also identical. For a proposal for the metaphysics of duplicates, in terms of natural properties, see Lewis (1983:~pp.~355-365).'} 

In conclusion: {\it two models are physically equivalent iff (i)~they are weakly theoretically equivalent, and (ii)~they have the} same {\it internal interpretation}: where `sameness' requires that an agreed philosophical conception of the interpretation has been supplied. This is the notion of theoretical equivalence that is suggested by dualities.\\

We can now be more precise about how weak theoretical equivalence can lead to physical equivalence, by comparing the diagrams in Figures \ref{dual3} and \ref{eqmapI}. Evidently, we obtain the diagram for physical equivalence, Figure \ref{dual3}, if the induced duality map, $\ti d$, in Figure \ref{eqmapI} happens to be the identity map (and provided the above-mentioned philosophical explication is also given). But this is not the generic situation, i.e.~$\ti d$ is in general {\it not} the identity map, as we saw in the example of Kramers-Wannier duality: since it maps high temperature to low temperature, and vice-versa. 

There is a second way to obtain a situation of theoretical, or physical, equivalence from weak theoretical equivalence, i.e.~Figure \ref{eqmapI}. The idea is to define a new pair of interpretations using the induced duality map, as follows: $i_1':=i_1$ and $i_2':=\ti d\circ i_2$, where both models now have $D_1$ as their domain of application, i.e.~$i_1'(m_1)=i_2'(m_2)=D_1$. This means that, in effect, we change $m_2$'s domain of application, from $D_2$ to $D_1$. (Although this definition treats the two models asymmetrically, there is an alternative definition that maps to $D_2$ rather than $D_1$). While this may at first sight look ad hoc, it is in fact natural if $i_2$ is an {\it external} interpretation, while $i_2'$ is now an {\it internal} interpretation:\footnote{If $i_2$ is external, then it maps the whole model, $M_2$, and not just the model triple $m_2$. In that case, the existence of an isomorphism $\ti d$ between the domains of application is very non-trivial, because the cardinalities of the models considered are no longer the same. However, it can be achieved if $i_2(m_2)=i_2(M_2)$, which in effect means that $i_2$ is already the external interpretation stripped of its ``external connotations'', i.e.~in effect restricted to the model root (and for $i_1'$ and $i_2'$ to both be internal interpretations that map to the same domain of application, the induced duality map $\ti d$ is not required to be an isomorphism).} and it is common practice among the physicists working on dualities. 

For example, consider gauge-gravity duality, where $m_1$ is a gauge theory model and $m_2$ is a quantum gravity model (more precisely, a version of string theory), and take $i_2$ be the string theory interpretation of the quantum gravity model, while $i_1$ is a sophisticated, internal interpretation of the gauge theory model. What we then learn about the interpretation of the quantum gravity model is that an {\it internal} interpretation, $i_2'$, of it exists: and this is natural, given the existence of a duality. This ``change of interpretation'', from the external interpretation $i_2$ to the new internal interpretation $i_2'$, in fact models the practices of the physicists who work on dualities: and the recent literature on dualities has tried to articulate in what sense such interpretations give natural interpretations of quantum gravity and quantum field theories. This is a large topic than I cannot discuss here, but it has been central to the recent literature on dualities: see, for example, Dieks et al.~(2015), De Haro (2016, 2019), Huggett (2017), De Haro and Butterfield (2018), Read and M\o ller-Nielsen (2018).\footnote{Recently, Weatherall (2019) has discussed a similar view in the context of classical electric-magnetic duality, where he introduces an `empirical significance functor' that relates the empirical substructures of the domains of application of the dual models. Weatherall, too, discusses the empirical {\it inequivalence} of electric-magnetic duals on their ordinary (in my language, `external') interpretations vs.~their empirical {\it equivalence} once an empirical significance functor is introduced (in my language, on an `internal' interpretation, restricted to the observational conditionals).}


\subsubsection{When is a judgment of physical equivalence justified?}\label{justify}

The above analysis of physical equivalence leaves unanswered an important epistemic question, about whether we are {\it justified} in adopting an internal interpretation, hence judging two models to be physically equivalent. 
The question is, roughly speaking, whether the internal interpretation is ``detailed enough'' and ``general enough'' that it cannot be expected to change, for a given domain of application $D$. I call this condition {\it unextendability,} and it is a condition on the interpreted theory. De Haro (2016: Section 1.3.3) gives a technical definition of the two conditions for unextendability, the `detail' and `generality' conditions. Roughly speaking, the condition that the interpreted theory is detailed enough means that it describes the entire domain of application $D$, i.e.~it does not leave out any
details. And the generality condition means that both the bare theory and the domain of application $D$ are general enough that the theory ``cannot be extended'' to a theory covering a larger set of phenomena. Thus the theory is as general as it can/should be, and $D$ is an entire possible world, and cannot be extended beyond it.\footnote{For some examples of unextendable theories, see De Haro (2016:~Section 2.3). Notice that, on my usage, a `model' actually describes an entire set of possible worlds (all the possible worlds that satisfy the dynamics). But it is simpler to phrase things in the jargon of `a possible world'.} For a discussion of unextendability in the context of the effective field theory programme, in De Haro (2016:~Section 1.4). 

Part of the unextendability requirement is a general point about the need for mathematically consistent models (which is not the same as: `mathematically rigorous models'), in order to address the question of when dual models may be taken to be physically equivalent. 
For example, most of the solutions of Newtonian gravitation run into inconsistencies. One can already see this in the most elementary examples: any two massive point particles placed next to each other will attract and run into each other, developing a singularity---an infinitely strong force, with the corresponding infinite potential energy---when the two objects coincide. And while this may be a consequence of idealizing massive bodies as point particles, all of the approaches that derive forces for extended bodies use, to my knowledge, the point particle case as their starting point (see for example Truesdell (1966:~Eq.~(1.4)), Arnold (1989:~pp.~133-135), Goldstein et al.~(2002:~pp.~134-136, 187, 224)). Thus the pathologies seem pervasive: and so, the theory---as currently formulated---is inconsistent: it does not make any predictions beyond a certain point.\footnote{For a review of the (related) problems associated with summing infinite numbers of point-particle forces acting on a massive body, and of the inconsistencies in cosmological applications of Newtonian gravitation, see Vickers (2013:~pp.~110-146). One might hope that there is a ``cleaned-up'' version of Newtonian gravitation that is free of such pathologies, while making the same claims about the world. But none of the strategies that I know of seem to work. For example, restricting the space of solutions to only those solutions that are regular would leave us with virtually nothing: certainly nothing of interest for the problem of two-body gravitational attraction that we started with, since it is from the two-body point particle case that we derive most other solutions, including the extended ones. Also, restricting the theory to the part of the domain of application where it is finite seems ad-hoc, and to not leave us with anything like a principled, suitably predictive theory for everyday objects---thus it is not clear that a domain of application exists on which the theory is finite. Finally, stabilising the singular solutions by including electrical repulsion not only changes the theory---we are then no longer dealing with Newtonian gravitation---but it also does not get rid of the inconsistencies, which now show up elsewhere. While we await whether a mathematically consistent amendment of Newtonian gravitation is forthcoming, it is fair to say that any such amendment is likely to give a {\it different} theory. } This may be no problem for ordinary uses of the theory, but it is a major obstacle if one wishes to discuss the putative physical equivalence between Newtonian gravitation and Newton-Cartan theory in a justified manner, as discussed above;\footnote{See Frisch (2005:~pp.~4, 9, 193-194), who defends the view that consistency is just one criterion of theory assessment, and that one can apply inconsistent theories provided that one supplies additional rules about how to apply the laws in various situations. In particular, he discusses causal constraints, considerations of simplicity and mathematical tractability, and rules guiding the use of different parts of the theory. I agree with this view, which admits that unextendable theories can be an exception (as Frisch (2005:~p.~8) himself admits, in the analogue case of a ``theory of everything'').} or between two versions of Newtonian mechanics with different standards of rest. For such discussions assume that the theory describes an entire physically possible world---which it does not!\footnote{Frisch (2005:~pp.~4, 7) also notices that consistency is built into both the semantic and the syntactic conceptions of theories and that any account of theories in terms of possible worlds assumes that the models (as instantiations of a theory) include structures `rich enough to represent possible worlds as complex as ours' (ibid, p.~8).}

The point is of course not new, and my critique is not a wholesale critique of possible world-talk in physics, as one might take e.g.~Wilson (2006:~pp.~201-203) to be (for other criticisms, see Frisch (2005:~pp.~4-9) and Vickers (2013:~pp.~243)).\footnote{The theme of the breakdown of a theory at a singularity as a ``smoking gun for new physics'' is of course well-known, as well: for example, it motivates K.~Wilson's approach to renormalization. But this does not conflict with my requirement of unextendable theories for a justified judgment physical equivalence, nor should that requirement be confused with an unreasonable demand that the theory be somehow ``final''. It is simply in the nature of physical equivalence that it requires mathematically well-defined theories.} Possible world-talk is helpful even when it is not meant literally---and so, it can be applied to physical equivalence if one is not worried about this judgment being justified. But if one wishes to justify the judgment of physical equivalence, for which all agree one needs ``a theory of the whole world'', one would expect that the possible world-talk should be taken to be literal. That is: because a justified judgment of physical equivalence (see above) entails unextendability, an entire possible world is required to be associated with the theory.

As I expounded in De Haro (2018:~Section 1.4), especially with the example of the effective field theory programme, it seems to me that claims of physical equivalence---but also any judgment of {\it theoretical} equivalence in physics---between incomplete or inconsistent theories are not justified. Namely, if one is dealing with an effective theory, one must always be prepared to see its interpretation change, precisely at those points at which the theory becomes inconsistent---and that may call for a major overhaul of the theory's interpretation. 
In such cases, talk of the `possible worlds' described by the theory cannot be literal, since there are no such worlds. At best, there are limited domains of application within larger worlds that are described by the theory. 
Thus general considerations of `toy cosmologies' or ``theories of everything'', without regard for the finiteness and consistency of the theory, are insufficient. Unextendability is a natural condition that is sufficient to fix this, cf.~De Haro (2016). I consider this to be one major contribution of dualities in string theory and quantum field theory to analyses of theoretical equivalence in physics.

\subsubsection{Comparing to other work on theoretical equivalence}\label{comothers}

In this Section, I comment on other recent work relating to theoretical equivalence, especially Butterfield (2018) and Read and M\o ller-Nielsen (2018). A common theme in the recent literature on dualities---and this is another aspect where dualities bear on discussions of theoretical equivalence in physics---is that ``physical equivalence is not automatic''. There is by now indeed a consensus among those working on dualities that duality does not invariably lead to physical equivalence (and I will substantiate the existence of this consensus in a moment).

My comments here aim to: (A) first, point out four points where these three authors went further than I did in De Haro (2015, 2016); and second, point out four specific interpretative points where the current paper adds to previous work: two are interpretative (B) and two are about sufficient conditions for physical equivalence (C). \\
\\
{\it A.~~~Four contributions of Butterfield (2018) and Read and M\o ller-Nielsen (2018)}\\
\\
The {\it novelty} of Butterfield (2018) and Read and M\o ller-Nielsen (2018) can be summarised in the following four points:

(1)~~While all hands agree that ``formally equivalent models can disagree'', i.e.~they can make different claims about the world even if their claims can be mapped to each other one-to-one, Butterfield (2018) has emphasised this point clearly and vividly, in several examples. 

(2)~~Furthermore, Butterfield (2018) {\it distinguishes two ways in which two models can disagree:} namely, by making contradictory assertions about the same subject-matter, or by describing different subject-matters. 

(3)~~Read and M\o ller-Nielsen (2018) have distinguished two different approaches to dualities, which they dub the `interpretational' and the `motivational' approaches (following a similar distinction, for symmetries, in M\o ller-Nielsen (2017)). By `approaches to dualities' I here mean different ways in which, when confronted with a duality, physicists or philosophers may choose to approach the interpretative project. The main difference is as follows: on the interpretational approach, the decision whether to interpret duality-related models as being physically equivalent {\it need not wait upon an explication of their shared ontology,} while on the motivational approach this explication is mandatory. (Within each of the two approaches, Read and M\o ller-Nielsen further distinguish two sub-positions, but I will not go into these details.) Thus theirs is a methodological distinction reflecting two different strategies for interpreting dual models.\footnote{Read and M\o ller-Nielsen (2018) and De Haro (2018) discuss complementary aspects of the methodology of dualities: while De Haro (2018) focusses on one aspect of the heuristic function of duality---namely, on  how duality functions within theory construction---Read and M\o ller-Nielsen (2018) focus on the closely related topic of approaches to interpretation---including, but not limited to, theory construction. Thus I disagree with Butterfield's (2018:~Section 1.2) characterisation, in his point (ii), of Read and M\o ller-Nielsen (2018) as being about the heuristic function of duality, rather than about interpreting dual theories as now formulated. For their analysis applies to both cases. In other words, their project is methodological but not always heuristic, in the sense of De Haro (2018).} Notice that the interpretational vs.~motivational contrast does {\it not} lead to new interpretative options regarding dual pairs: even though being an interpretationalist or a motivationalist does of course bear on the kinds of interpretations that one is likely to adopt. 

(4)~~ Furthermore, Read and M\o ller-Nielsen (2018) articulate the possibility that {\it dual models might not have physical interpretations,} and they give this possibility a place in their overall methodological account---namely, in what they call the `cautious motivational view' 
(more on this below, especially in footnote \ref{RMN}: and, as should be clear already from Section \ref{comptheq}, I agree with these authors about motivationalism being the preferred position).\footnote{Extendable theories provide good examples of Read and M\o ller-Nielsen's (2018) observation, in point (4), that an explication of the shared ontology of two models need not exist (see the discussion in the previous Section). For inconsistent models only have limited ontologies, and so the question of the justification of physical equivalence does not arise---there is no possible world described by the theory. However, Read and M\o ller-Nielsen's (2018) observation (4) does not amount to the theory's unextendability: for the former is a purely interpretative condition, while unextendability is both formal and interpretative.\label{RMN}}\\ 
 \\
{\it B.~~~Interpretative attitudes and quietism}\\
\\
Let me now make a friendly mention of two points where I think that, in light of the present work, the two discussed accounts of physical equivalence are not enough. The first leads into the second: the first is about differing conceptions of physical equivalence in the literature; the second is about quietism and realism.

The interpretational vs.~motivational contrast of Read and M\o ller-Nielsen (2018) gives an interesting characterisation of different approaches to interpretation. The main contrast is that motivationalists say we are warranted to pronounce verdicts about physical equivalence only if an explication of the common ontology of the models has been given. But, in so far as it is a contrast based on the {\it actions} or {\it attitudes} taken by interpreters of dualities, rather than on the {\it reasons} leading to those actions or attitudes, or on the {\it notion of physical equivalence} itself, they do not seem to amount to fully fleshed-out intepretational stances. In particular, interpretationalism is heterogeneous and does not seem to represent a single interpretative position. Their common trait is that interpretationalists say we are warranted to draw verdicts of physical equivalence even in the absence of an explication of the common ontology. 

There is a third approach in the recent literature that, I think, is much more widely represented than interpretationalism, which in Section \ref{gdte} I introduced as {\it quietism}: namely, the stance of the authors that I quoted in Section \ref{gdte} as the ``consensus'' regarding theoretical equivalence, who are engaged with the formal project of theoretical equivalence, and for whom interpretative equivalence is a minimal requirement that their project needs, but on which they do not wish to focus. 

Interpretationalism seems to be less represented in the literature than one might at first think. (Dewar (2015) seems to be one.) For example, Read and M\o ller-Nielsen (2018:~Section 5.1) quote Rickles as an example, but I do not see that the evidence for his interpretationalism is more compelling than that for motivationalism or, indeed, quietism. One reason for this is, I think, that Rickles (2017:~p.~64, my emphasis) is not committed to a realist notion of physical equivalence: `the equivalence is at the level of theories first and foremost... We must therefore remain {\it purely in the realm of theoretical descriptions} to avoid such `external' [physical] confounders.' Thus it seems to me that Rickles is best seen as a quietist.

Read and M\o ller-Nielsen (2018) assume realistic interpretations of scientific theories. Also Butterfield (2018) endorses realism in the sense of referential semantics. But I think that referential semantics does not settle all ontological matters because not every ontology, in the sense of ``realist'' referential semantics, is realist in the metaphysical sense (see the discussion in Section \ref{seminter}). Therefore, I have kept metaphysics as far from the main substance of my proposal for theoretical equivalence as possible---specific metaphysical discussions only coming in when we discuss physical equivalence. For the theory's interpretation admits both realist and non-realist explications and different resolutions of referential indeterminacy (cf.~Section \ref{comptheq}). \\
\\
{\it C.~~~Sufficient conditions for physical equivalence and consistency}\\
\\
It is worth mentioning two further points where I think that, in light of the present work, the accounts of physical equivalence in the literature are not enough. The first is about sufficient conditions for equivalence; the second is about the mathematical consistency required for physical equivalence.


First, Butterfield (2018) does not list additional requirements on the kinds of bare theories or of interpretations that, given a case of duality, can lead to physical equivalence. For example, Butterfield (2018) is sympathetic to my conditions of unextendability and internal interpretation, and he even uses the question of whether the theories in question are `toy cosmologies' or ``theories of everything'' (TOEs) as a criterion to determine whether they are about the same subject-matter (2018:~\S4.1-\S4.5, \S6.2). The reason seems to be `that for two dual TOEs, there is literally `no room in the cosmos' for them to have separate but isomorphic subject-matters' (\S1.2). However, he does not address the question of sufficient conditions for taking duality to give physical equivalence.\\

Second, so far as I can see, the mathematical inconsistencies of some of the theories that are taken as examples of duality and theoretical equivalence (inconsistencies which I discussed in Section \ref{comptheq}) have been largely overlooked in the justification of judgments of theoretical equivalence: for example, in Newtonian gravitation. At least, it has not voiced the problem as clearly as it could have. For example, Read and M\o ller-Nielsen (2018:~Section 3.2) take Newtonian gravitation as their main detailed example.\footnote{I will be guilty of the same sin in the next Section: Maxwell's theory, as usually formulated, is not unextendable.} And, even though this example aims to illustrate {\it symmetry}, and only by {\it analogy} duality, the authors `seek to provide a fully worked out example of what it {\it means} to fully explicate symmetry-related models' underlying ontology' (ibid.). But, as I discussed in Section \ref{comptheq}, I doubt that one can fully explicate the ontology of Newton's theory of gravitation without addressing its singularities and-or inconsistencies. Also Butterfield (2018:~Section 4) discusses various examples of dualities: Newtonian mechanics with different standards of rest, position-momentum duality in quantum mechanics, Kramers-Wannier duality, Heisenberg vs.~Schr\"odinger pictures in quantum mechanics, and Legendre vs.~Hamiltonian classical mechanics. Agreed: mathematical consistency is less central to Butterfield's analysis, since he does not aim at a full explication of the ontology of these theories. But the fact that also Butterfield does not distinguish between theories that are well-defined (like his example of position-momentum duality in quantum mechanics) and those that have singularities (like his example of Newtonian gravitation) confirms my point. \\

In Section \ref{comptheq}, I argued that one must have consistent models in order to regard them as justified cases of theoretical equivalence. This also helps one overcome the rhetoric of ``examples of dualities should not depend on advanced theoretical physics''. 
While the simplicity of the examples is desirable, it cannot be had through inconsistent theories, if what one wishes to discuss is the question of when one is justified in judging cases of theoretical equivalence---which forces us to consider consistent theories.\footnote{De Haro and Butterfield (2018:~Sections 4-5) contain what I think is the first (and perhaps the simplest) rigorous example of a duality, for a quantum field theory.} Indeed, dualities between inconsistent models such as Newtonian gravitation are {\it not} justified cases of theoretical equivalence, and so they can at best be useful as pedagogical `toy models'---as Butterfield indeed seems to take them. In a slogan: a contemporary discussion of theoretical equivalence must go beyond Leibniz, Clarke, and Galileo's ship. 

It seems that considerations of physical equivalence---and more specifically, the criterion of unextendability---give us an interesting distinction in the types of theories that are subject to duality: those in which there is an interesting question about the justification of the judgment of physical equivalence under all suitable interpretations, and those for which this question is answered in the negative.

\section{Maxwell's Theory of Electromagnetism}\label{Maxem}

This Section contains the case study illustrating the Schema and the notion of theoretical equivalence proposed in the previous Section: namely, Maxwell's theory of electromagnetism. 
Sections \ref{EMT} and \ref{emm} introduce the {\it bare} theory and models, respectively. Section \ref{teMx} explores theoretical equivalence, and so an interpretation is only required in that Section. However, already in Sections \ref{EMT} and \ref{emm} I will use the language of interpretation, in the text around the formulas. This will be the standard theoretical physics talk of a `Faraday tensor', `Maxwell's equations', a `gauge field', etc.~(and these terms already assume an at least minimal physical interpretation), rather than the abstract talk of a `rank-2 antisymmetric tensor', a `linear inhomogeneous second-order differential equation', a `Lie-algebra-valued one-form', etc., which would be both tedious and unnecessary---since we will need the physics talk in Section \ref{teMx}. This then reflects physical practice, where bare theories are usually already, at least partly, interpreted. But such practice does not commit us to a specific interpretation. For the same reasons, I will use the terms `theory' and `models' rather than `bare theory' and `bare models'.

Also, since we will only consider internal interpretations, I will occasionally say that the model root `contains the physically significant' parts of a model, and so on---which is indeed appropriate on an internal interpretation.

My choice of case study is partly motivated by the aim of being able to compare, in upcoming work, the Schema's judgments with Weatherall's (2015, 2016, 2016a) proposal of categorical equivalence as a criterion for theoretical equivalence (see the discussion in the Introduction): and of course by the fact that Maxwell's theory is an important and well-known example in physics.

\subsection{The electromagnetic theory}\label{EMT}

I begin, in this Section, by casting classical Maxwell's theory in the language of the Schema. 

In a classical theory, a state is specified by a field or other variable obeying appropriate requirements, for example linearity (a linear superposition of two fields again gives a field). At this stage we want to stay general, and so we do not require that a field solve the equations of motion---the Lagrangian and Hamiltonian formalisms, for example, require the use of fields that do not satisfy the equations of motion. Thus we will take the state-space of Maxwell's theory of electromagnetism to be specified by elements that are quadruples, $({\cal M},g,F,J)$, of: (i) a manifold, ${\cal M}$ (here, 4-dimensional: and, in fact, $\mathbb{R}^4$); (ii) an invertible, smooth flat metric, $g$, with Lorentzian signature; (iii) a Faraday tensor, $F$ (i.e.~a square-integrable, rank-2 antisymmetric tensor);\footnote{Thus Faraday tensors are members of the following space: $F\in{\cal F}:= L^2(\L^2({\cal M},\mathbb{R}))$.} (iv) a conserved vector current, $J$, with compact support.\footnote{Much of the literature tends to not consider conditions such as the square-integrability of the Faraday tensor and the compact support of the currents: but some such condition is required in order for the quantities to be well-defined. Although square-integrability and compact support are strong conditions which can be relaxed (the relaxation giving rise to a host of interesting and important questions such as the status of ``large gauge transformations''), we will for simplicity not consider such weakenings here.\label{boundc}
} Thus, the state-space is a quadruple, ${\cal S}=({\cal M},g,{\cal F},{\cal J})$, of a fixed manifold and metric, a set of antisymmetric 2-forms (Faraday tensors), and a set of 1-forms (conserved currents). Thus a state in this theory assigns a Faraday tensor throughout all of ${\cal M}$, not just an instantaneous state. 

I will denote the members of the sets of Faraday tensors and conserved currents, respectively ${\cal F}$ and ${\cal J}$, by $F_i$ and $J_i$, where $i\in I$ for some index set $I$. 
And it will be useful to abbreviate the state-space and denote it by two of its members, i.e.~the pair $({\cal F},{\cal J})$, and even just by the set of Faraday tensors, ${\cal F}$. (I will sometimes also, slightly abusing notation, write e.g.~$F\in{\cal S}$ or $J\in{\cal S}$, which is of course short for: $F\in{\cal F}$, with ${\cal F}$ the third component of the state-space, ${\cal S}$, and $J\in{\cal J}$, with ${\cal J}$ the fourth component of the state-space.) Notice that we have not yet specified that the Faraday tensors and the conserved currents must be solutions of the equations of motion.\footnote{This is what physicists call working ``off-shell'', and it is needed e.g.~for discussing the Lagrangian and Hamiltonian formalisms.\label{offshell}} The symmetries, $a$, of the state-space are the diffeomorphisms (usually, for a flat metric: Poincar\'e transformations) that preserve the metric, $g$.

But this is of course not sufficient to define a theory. We also need to specify the quantities that can appear in the theory, i.e.~the set ${\cal Q}$. Recall, from Section \ref{thmod}, that it is useful to think of quantities as being the duals (in the mathematical sense!) of states, i.e.~as maps $Q:{\cal S}\rightarrow\mathbb{R}$. At the linear level, we have a map whose image is the Faraday tensor itself, evaluated at a point $p\in{\cal M}$, i.e.~$Q_1[F]=F(x)$~. At the quadratic level, a distinguished quantity in electrodynamics is the stress-energy tensor: it takes a Faraday tensor as input, and it outputs the value of a quadratic polynomial in the Faraday tensor (cf.~also Section \ref{seminter}). Its components are: 
\bea\label{Tmn}
T_{\m\n}[F] =-{1\over\m_0}\,F_{\m\l}(x)\,F^\l{}_\n(x)-{1\over4\m_0}\,\eta_{\m\n}\,F_{\l\s}(x)\,F^{\l\s}(x)~,
\eea
where $\m_0$ is the vacuum permeability. The 00-component is of course interpreted as the electromagnetic energy (cf.~Eq.~\eq{emE}). Thus we can (at least tentatively: see the further development below) regard a quantity, $Q[(g,g^{-1},F,J)]$ as a functional of the fields, including the metric and its inverse. Since we want to consider a local, covariant theory, we will take the values of the quantities to be any covariant, local polynomials (evaluated at the same spacetime point) that can be constructed from tensor powers of the metric, the Faraday tensor, and the current, i.e.~any polynomials in the fields $g,g^{-1},F,J$, with any covariant contractions of their indices. 

However, we need a more general set of quantities, since integrals over suitable regions, and derivatives of these polynomial quantities, calculated using the $g$-compatible connection, are also admissible, as long as they are covariant. For example, a typical quantity is quadratic in the Faraday tensor, viz.~$\int_{\cal M}F\wedge*F$, which gives the Lagrangian of electrodynamics.\footnote{Here are some details, which we will not need to use explicitly. We can include differentiation and integration operators that act on the quantities. Thus the set of quantities, ${\cal Q}$, is a set of maps, whose images form an (abelian) algebra: namely, the universal enveloping algebra (i.e.~the set of all local polynomials) over the reals, generated by $g,g^{-1},F,J$ and by the derivative and integral operations acting on them, which I will denote by $\dd$ and $\int$. Notice that, by construction, the quantities are covariant under the symmetries. I will denote by $Q\left(g,g^{-1},F,J;\dd,\int\right)$ the universal enveloping algebra of these fixed elements, $g,g^{-1},F,J;\dd,\int$, just discussed.} By a slight abuse of language, I will speak of the images of the maps, rather than the maps themselves, as being the `quantities'. The total set of quantities (their images) then includes the action of the differential and integral operators, $\dd$ and $\int$, on the fields: I will denote it by ${\cal Q}:=\{Q\left(g,g^{-1},F,J;\dd,\int\right)\}_{F\in{\cal F},\,J\in{\cal J}}$.\footnote{Unlike $Q\left(g,g^{-1},F,J;\dd,\int\right)$, which for fixed $F$ is both additive and multiplicative, ${\cal Q}$ is not (for example, we cannot add or multiply terms with different $F$'s, except in the linear quantities).}

The dynamics is again a structured set, consisting of two equations (assumming a flat metric in the second equation):\footnote{Notice that there is some flexibility as to where to put the Bianchi identity, i.e.~the first of Eq.~\eq{Mxeq}: it could as well have been taken to be part of the definition of the state-space, ${\cal S}$, rather than the dynamics, ${\cal D}$. In that case, we would have restricted the whole discussion to closed 2-forms. But, in view of possible generalisations of this example that include magnetic monopoles, it is best to keep the Bianchi identity as part of the dynamics (and this is also the normal physical usage, i.e.~the Bianchi identity is part of Maxwell's equations, namely it is the 2-form field version of Gauss's law for magnetism and Faraday's law).}
\bea\label{Mxeq}
{\cal D}:=\left\{(F,J)\in\cal{S}~\Big|\begin{array}{ccc}
~~~\dd F&=&0\\
\dd*F&=&\m_0\,J
\end{array}\right\}~.
\eea
This set is a linear vector space. Namely, take pairs $(F_1,J_1),(F_2,J_2)\in{\cal S}$, each satisfying Eq.~\eq{Mxeq}: then, under linear addition, the resulting pair, $(F',J'):=(a\,F_1+b\,F_2, a\,J_1+b\,J_2)\in{\cal S}$, which is also be a member of the state-space (with $a,b\in\mathbb{R}$), also satisfies the dynamics, Eq.~\eq{Mxeq}. In short: $F'$ and $J'$ also satisfy Eq.~\eq{Mxeq}. The symmetries of course leave the dynamics invariant, as they shold.

Notice that I have made the choice that ``the theory is off-shell'' while in the next Section we will take the models to be on-shell: ``the theory being off-shell'' is physics jargon for the theory's fields not being required to satisfy the dynamics, while ``the models being on-shell'' is jargon for the models' fields satisfying the dynamics (i.e.~the set of dynamical equations), Eq.~\eq{Mxeq}.\footnote{The reason for taking the theory to be off-shell, as mentioned in footnote \ref{offshell}, is that this is required for the Lagrangian and Hamiltonian formalisms. On the other hand, I will take the models to be on-shell because this choice will give us, in upcoming work, a way to embed Weatherall's categories into the Schema. Notice that Weatherall's theories are our models, and that he does not discuss theories in the Schema's sense of a bare theory (that is represented homomorphically by models in our sense).} This choice is fine: the representation map from the theory to the model triples then maps off-shell fields and quantities to on-shell ones, and it maps dynamical equations that impose a constraint to dynamical equations that are identically satisfied: and so, the representation map is a homomorphism (see the next three Sections).

What are the models of this theory, i.e.~its representations? Let me now discuss some fairly straightforward representations of $T$ that we can construct, and which correspond to the models that Weatherall (2015) discusses (see the last paragraph in the preamble to this Section). 

\subsection{The electromagnetic models}\label{emm}

In this Section, I introduce the three models, whose theoretical equivalence we will assess. Section \ref{Fm} introduces the Faraday model. Section \ref{gfm} introduces the gauge field model. Finally, section \ref{gom} introduces the gauge orbit model. 

\subsubsection{The Faraday model}\label{Fm}

The Faraday model, which I introduce in this Section, is directly based on the formulation of the theory, $T$, discussed in Section \ref{EMT}. For there is of course a representation of $T$, i.e.~a model in my sense from Section \ref{thmod}, that is isomorphic to the subset of the theory satisfying the dynamics: namely, the state-space is a set of quadruples, $({\cal M},\eta,F,J)$, satisfying the equations of motion, Eq.~\eq{Mxeq} (and from now on, we take ${\cal M}=\mathbb{R}^4$ and for simplicity I will fix the metric $g$ to be the Minkowski metric, denoted by $\eta$). This set consists of all the solutions $(F,J)$ to the equations of motion, and they are distinguished by their boundary and initial conditions and symmetries. 
We can write this as follows:
\bea
{\cal S}_{\tn{Faraday}}:=\{\,({\cal M},g,F,J)~|~(F,J)\in{\cal D}_{\tn{Faraday}}\,\}~,
\eea
where the elements of the quadruple are defined as above, and the dynamics, ${\cal D}_{\tn{Faraday}}$, is the set of pairs $(F,J)$ satisfying the two equations in Eq.~\eq{Mxeq}. Again, the above set of quadruples comes with a structure, viz.~the model is closed under addition (i.e.~the sum of two solutions is again a solution). The set of quantities, then, is the set of functionals on the state-space, whose images are constructed from the polynomials in the fields, as before: ${\cal Q}_{\tn{Faraday}}:=\{Q(g,g^{-1},F,J;\dd,\int)\}_{(F,J)\in\,{\cal S}_{\tn{Faraday}}}$. I shall denote this model triple by $m_{\tn{Faraday}}$. 

\subsubsection{The gauge field model}\label{gfm}

This Section introduces the second model, namely the gauge field model: which we will denote as $M_{\tn{GF}}$.  This model is obtained from the observation that the closure condition of the Faraday tensor, i.e.~the first of Eq.~\eq{Mxeq}, can be solved (if ${\cal M}=\mathbb{R}^4$) by introducing a smooth 1-form, $A$, such that:
\bea\label{fda}
F=\dd A~.
\eea
If $A$ is smooth, then identically also: $\dd F=\dd^2A=0$. In addition, $A$ satisfies $\dd*\dd A=\m_0\,J$. The state-space of the model then consists of the quadruples $({\cal M},\eta,{\cal A},{\cal J})$, viz. ${\cal S}_{\tn{GF}}=\{({\cal M},\eta,A,J)|_{\sm d*\sm d A=\m_0J}\}$. 

In this model, something interesting and important happens: namely, the quantities may now depend on $A$. Since $A$ is part of the specification of a state of a model, the set of quantities, ${\cal Q}_{\tn{GF}}\left(\eta,A,J;\dd,\int\right)$, is the set of maps on the state-space whose images are the covariant polynomials in $A$ and $J$, together with their derivatives and integrals. This is because of the general relation between state-space and quantities for classical field theories that we discussed in Section \ref{EMT}: and it contrasts with the previous model, where the quantities were restricted to those built from $F=\dd A$ and $J$ only, i.e.~no direct $A$-dependence was allowed (in fact, no gauge field was defined). 

To view this as a representation of the theory, $T$, we have to construct the representation map from ${\cal S}$ to ${\cal S}_{\tn{GF}}$ (see Eq.~\eq{hmtm}). In this representation we assign, to each Faraday tensor, $F\in{\cal S}$, an image gauge field, $A\in{\cal S}_{\tn{GF}}$:
\bea\label{hrf}
h^{\tn{GF}}(F)=A~.
\eea
However, this map is of course not unique. As is well known, for any smooth scalar function $\l$, the vector potentials $A$ and $A':=A+\dd\l$ give rise to the same Faraday tensor. This means that, for each smooth $\l$, there is a different map Eq.~\eq{hrf}. But this just means that we have different maps $h^{\tn{GF}}$, each mapping to $A+\dd\l$ for different smooth functions $\l$ (Eq.~\eq{hrf} being the $\l=0$ case). 

In fact, it is easy to show that each $A_i$, $i\in I$ (the image of $F_i\in{\cal S}$ under the map Eq.~\eq{hrf}, where $I$ is an index set labelling all solutions of Maxwell's equations) must have its own smooth function $\l_i$, so that the correct definition of the homomorphism is:\footnote{The homomorphism $h^{\tn{GF}}$ is linear, because ${\cal S}$ is a linear space. Now if each map comes with {\it the same} smooth function $\l$, the map is not linear: $h^{\tn{GF}}(F_1+F_2)=h^{\tn{GF}}(\dd A_1+\dd A_2)=h^{\tn{GF}}(\dd(A_1+A_2))=A_1+A_2+\dd\l\not=h^{\tn{GF}}(A_1)+h^{\tn{GF}}(A_2)$. This is remedied if $h^{\tn{GF}}$ is defined as in Eq.~\eq{hGH}.}
\bea\label{hGH}
h^{\tn{GF}}(F_i)=A_i+\dd\l_i\,\in\,{\cal S}_{\tn{GF}}~~,~~~~i\in I~.
\eea
Thus, as far as the gauge field is concerned (suppressing in the notation all the other parts of the states, and also the quantities) a model state-space looks like: 
\bea
{\cal S}_{\tn{GF}}=\Big\{A_i+\dd\l_i~\Big|~i\in I~,~\dd*\dd A_i=\m_0\,J_i\Big\}~.
\eea
And likewise when we include quantities, $\{Q^{\tn{GF}}(\eta,A,\l,J)\}$, and dynamics.\footnote{This model, once internally interpreted, is indeterministic: for a specification of initial conditions for a state is compatible with infinitely many solutions in the future, the indeterminism being parametrised by the gauge parameter $\l$ (cf.~Belot (1995:~pp.~88-90)). Also, the quantities $Q$ that depend on the gauge field $A$, rather than on the Faraday tensor, are multi-valued. But nevermind: my aim in this Section is not to give a familiar model, but rather to illustrate two different ways of defining the model triple (by contrasting this Section and the next).}

Rather than having a single model, we now have a model for each set of gauge functions $\{\l_i\}_{i\in I}$. We will denote the different models by $m^{\tn{GF}}_\l$. 

Since $m^{\tn{GF}}_\l$ keeps $A$ in its space of states, with a corresponding exact piece $\dd\l$ chosen by the stipulation, Eq.~\eq{hGH}, the map $h^{\tn{GF}}$ induces a symmetry of the state-space, ${\cal S}_{\tn{GF}}$. Namely, it induces a translation along the gauge orbits which are images of the same value of the Faraday tensor. So, ${\cal S}_{\tn{GF}}$ comes equipped with an additional set of symmetries: let us dub this set of symmetries $\s_{\{\l_i\}}$.

But it is important to note that in this model, we do not define equivalence classes for these symmetries, i.e.~there is no judgment that  gauge fields need to be identified along gauge orbits: in fact, in this model they are {\it not} identified. 
Notice that such a model, where purportedly ``gauge'' structure is part of the triple---and hence, on an internal interpretation, is physical---{\it is} a representation of $T$, since Eq.~\eq{hGH} is a homomorphism.


\subsubsection{The gauge invariant model}\label{gom}

This Section introduces a third model, the gauge invariant model, which is immediately prompted by the discussion at the end of the previous Section. I will denote it by $m_{\tn{GI}}$. 

In this model, we 
take the homomorphism for the state-spaces to map to the set of gauge orbits with representative $A_i$:
\bea\label{gifa}
h_{\tn{GI}}(F_i)=[A_i]:=\{\,A_i\sim A_i+\dd\l~|~\l:\mathbb{R}^4\rightarrow\mathbb{R}\,\}\,,~~~i\in I~.
\eea
All the gauge fields related by an exact one-form belong to a gauge orbit that passes through the (arbitrarily chosen) representative of the orbit, $A_i$, and they are identified under the equivalence relation.\footnote{Presented with such models, where the state-space is quotented by a symmetry, there is a natural question of whether it is mandatory to move to a formalism that automatically enforces the symmetry. Dewar (2018) has for example introduced a jargon distinguishing a `reduced theory', where the symmetry is enforced by the formalism, and a `sophisticated' version of the theory, where the formalism does not enforce the symmetry---symmetries then act e.g.~as isomorphisms. Dewar (2018) advocates the sophisticated version: and I agree that a reformulation of the theory that enforces the symmetry is not mandatory, even if it can be useful. The present formulation of Maxwell's theory is in terms of the gauge dependent variables, and the symmetry is implemented through an identification. This is reflected in the use of specific structure, as I show below, which on an internal interpretation will indeed be redundant.}

Thus the space of one-forms of this model---which is the central part of the state-space---is the quotient of the state-space of the previous model by the gauge symmetry: namely, it is the set of gauge orbits that satisfy Maxwell's equations:
\bea\label{ago}
{\cal S}_{\tn{GI}}&:=&{\cal S}_{\tn{GF}}/\!\sim\,\,
=\{\,[A_i]~|~i\in I~,~\dd*\dd A_i=\m_0\,J_i\,\}~.
\eea
Thus the states depend only on the gauge equivalence class. Notice that, in the present case of classical electromagnetism on ${\cal M}=\mathbb{R}^4$ with a flat Minkowski metric, specifying an equivalence class $[A_i]$ is equivalent to specifying a Faraday tensor $F_i$. But the construction of the state-space uses specific structure: both in the choice of a representative of the gauge orbit $A_i$ (and Maxwell's equations $\dd*\dd A_i=\m_0\,J_i$ are written down using this representative) and in the choice of gauge functions $\l$ that move one along the gauge orbit.

Thus the full state-space of this model is given by all quadruples $({\cal M},\eta,[A],J)$, satisfying the equations of motion. The quantities of this model are the same as in the Faraday model, but written in terms of the gauge field $A_i$ rather than in terms of the Faraday tensor.\footnote{Recall, from the previous Section, that quantities that depend directly on $A$, rather than through the Faraday tensor, were multi-valued. In this model, we now drop such quantities, because they are incompatible with the equivalence relation, Eq.~\eq{gifa}, on the state-space . 
} The full model then is: $M_{\tn{GI}}=\bra m_{\tn{GI}}\,;\,\l\ket$, where $A$ is an arbitrary representative of the gauge-orbit that enters in the definition of the state-space, Eq.~\eq{ago}, and $\l$ in the definition of the equivalence class, Eq.~\eq{gifa}.

In this case, the homomorphism from the theory to the model is unique, and given by Eq.~\eq{hrf}. This is because the gauge field $A$ is defined only up to an exact one-form.

\subsection{Theoretical equivalence in electromagnetism}\label{teMx}

In this Section, I collect our findings from the previous Section regarding our model triples, $m_{\tn{Faraday}}$, $m_{\tn{GF}}$, and $m_{\tn{GI}}$, so as to assess theoretical equivalence, according to the conditions (A') and (B') in Section \ref{gdte}.\\

I first discuss (A'), i.e.~the isomorphism of model roots (here, model triples). Recall that the state-space of the Faraday model triple, $m_{\tn{Faraday}}$, is the space of quadruples $({\cal M},\eta,F,J)$ satisfying Maxwell's equations. On the other hand, the state-space of the gauge invariant model triple, $m_{\tn{GI}}$, consists of the quadruples $({\cal M},\eta,[A],J)$, satisfying Maxwell's equations. And we saw that the gauge orbits, $[A_i]$, are in one-to-one correspondence to the Faraday tensors, $F_i$, because each gauge orbit fixes a Faraday tensor, and vice versa. And also the dynamics between the two models is the same, though written in different variables. So, to see if the spaces are isomorphic, we only need to see whether the quantities defined on the phase space are isomorphic: and they are, for they are the same set of quantities, only written either in terms of the Faraday tensor or in terms of the gauge field. Thus we conclude that the Faraday and the gauge invariant model triples are isomorphic, $m_{\tn{Faraday}}\cong m_{\tn{GI}}$. So to check their theoretical equivalence, we also need to check whether they satisfy condition (B'), i.e.~if their internal interpretations match (cf.~below). 

On the other hand, $m_{\tn{Faraday}}$ and $m_{\tn{GI}}$ are {\it not} isomorphic to any of the models $m_{\tn{GF}}$, because Faraday tensors are {\it not} in bijection with gauge fields $A+\dd\l$, as indicated by the existence of multiple homomorphisms, Eq.~\eq{hGH} (since this model does not quotient out the gauge symmetry). Furthermore, the gauge field model triples, $m_{\tn{GF}}$, contain strictly more quantities than $m_{\tn{Faraday}}$ and $m_{\tn{GF}}$, since the powers of $A$ are also quantities of $m_{\tn{GF}}$.\footnote{Of course, the quantities of $m_{\tn{GI}}$ {\it can} be obtained from the quantities of $m_{\tn{GF}}$ by projecting out all polynomials in $A$. But anyhow, their state-spaces are not isomorphic.} 


Thus, since $m_{\tn{Faraday}}$ and $m_{\tn{GF}}$ are not isomorphic, they cannot be theoretically equivalent on their internal interpretations. 

I now return to (B') (cf.~Section \ref{sdc}) for $m_{\tn{Faraday}}$ and $m_{\tn{GI}}$. This is the question whether there is an induced duality map, $\ti d$, between the domains of application of the two model roots, $m_{\tn{Faraday}}$ and $m_{\tn{GI}}$. We only need to check this for the state-space, since the quantities and dynamics trivially map onto each other. And we only need to check this for the Faraday tensor and the gauge field, since the three remaining components, ${\cal M}, \eta,J$, of the state-space are the same.

The interpretation of the state-space in the Faraday model, $m_{\tn{Faraday}}$, is as follows (cf.~Eq.~\eq{i(F)}):
\bea\label{iFar}
i_{\tn{Faraday}}(F)&=&\mbox{`a coordinate-free specification of an electric and magnetic configuration,}\nn
&&~~~~~~~~~~~~~~~~~~~~~~~~~~~~~\mbox{from a source $J$'}~.
\eea
As in Section \ref{seminter}, this interpretation is an intension: and it assumes that the (non-dynamical) `source $J$' has an established meaning, often by abstraction, through experimentation and instrumentation. An extension would typically specify additional details about the source, and how it is manipulated in experiments.

In the gauge invariant model, $m_{\tn{GI}}$, we interpret the state-space as follows:
\bea\label{iGI}
i_{\tn{GI}}([A])&=&\mbox{`a coordinate-free specification of a state of light from a source $J$'.}
\eea
Again, this intension assumes that `a state of light' has an established meaning---for example, in terms of the direction of motion, wave number, and polarizations of light waves---which together again specify electric and magnetic configurations, from the source $J$.

Since each gauge orbit $[A]$ uniquely specifies a Faraday tensor and vice versa, the following induced duality map is well-defined as an isomorphism:
\bea\label{asFt}
\ti d\,(\mbox{`an electric and magnetic configuration'}):=\mbox{`a state of light'}.
\eea
Thus the two models are {\it weakly theoretically equivalent}: in one model, the states are the states of the electric and magnetic fields, while in the other they are states of the (gauge invariant part of the) gauge field, i.e.~a vector field. 

But I want to argue that we should draw a stronger conclusion: namely, that the induced duality map is the identity map, $\ti d=\mbox{id}$, so that the models are {\it theoretically equivalent}, relative to these particular interpretations (cf.~Section \ref{comptheq}). Indeed, the two interpretations simply seem to respond to different historical perspectives on the same phenomenon (electric and magnetic phenomena vs.~phenomena involving light). And Maxwell's unification (synthesising the insights and efforts of Faraday, \O rsted, and many others) showed that electromagnetic and light phenomena were {\it the same}. Thus, $[A]$ and $F$ are completely equivalent in the bare model, and therefore their internal interpretations are the same, i.e.~they correspond to the same items in the world (the same current and the same electric and magnetic fields). This is an identification of physical phenomena, prompted by experiment, which now has become part of how the theory is understood;\footnote{It is also allowed by the formalism: for recall that, in the bare model triple $m_{\tn{GI}}$, a gauge orbit $[A]$ corresponds to a unique Faraday tensor $F$, and vice versa. Although mathematically these are not the same objects, their function within the theory is the same---a specification of a state in the state-space as a gauge orbit is the same as a specification as a Faraday tensor. (They also lead to the same quantities.)} it is not an additional formal requirement. Thus:
\bea
\mbox{`electric and magnetic configurations'}=\mbox{`a state of light'}~.
\eea
Using this, Eq.~\eq{asFt} simply becomes an identity, and the two models are physically equivalent, relative to this interpretation. That is: we have physical equivalence in the sense of Figure \ref{dual3}.

\section{Conclusion}\label{conclusion}

In this paper I have given an explication of theoretical equivalence based on  the Schema for theories and duality, from De Haro (2016) and De Haro and Butterfield (2017). The Schema's account proposes interpretative and formal conditions of theoretical equivalence, in two steps. In the first step, i.e.~{\it weak theoretical equivalence,} the interpretative condition is the requirement that two theoretically equivalent models should have matching internal interpretations, i.e.~isomorphic domains of application. The formal condition is isomorphism of models, but of a special kind: viz.~isomorphism between model roots that are homomorphs of a single theory. More specifically, when the models are triples, we have {\it triples} of isomorphisms. In the second step, i.e.~{\it physical equivalence,} the domains of application are required to be the same.

The question of when one is justified in making a judgment of physical equivalence can be answered by requiring that the models are unextendable. This suggests that 
many of the toy examples of duality and of theoretical equivalence in the literature (including the example discussed in Section \ref{Maxem}) are not justified cases of physical equivalence---and so, the completion of these models may well modify such judgments. Examples of dualities with unextendable theories are in De Haro (2016), Huggett (2017), and De Haro and Butterfield (2018). 

Elsewhere (De Haro (2016) and De Haro and Butterfield (2017)) we showed that the Schema gives the right judgments about bosonization and about (a simple example of) gauge-gravity duality. Work is underway to apply it to other examples. 

Thus the Schema is seen to give an interesting and plausible notion of theoretical equivalence, which can be applied to a broad class of cases---including cases, discussed in this paper, for which it was not designed.

This also suggests that several of the interpretative questions that have been discussed in detail not just in this paper, but by other recent authors working on dualities---about internal vs.~external interpretations, about unextendability and the justification of physical equivalence, and about the philosophical explication of interpretations\footnote{For some references where these interpretative questions are studied in detail, see Dieks et al.~(2015), Huggett (2017), De Haro (2015, 2016, 2019), Read and M\o ller-Nielsen (2018), De Haro and Butterfield (2017), Butterfield (2018).}---should be of interest for accounts of theoretical equivalence that aim to apply to physical theories.

\section*{Acknowledgements}
\addcontentsline{toc}{section}{Acknowledgements}

I thank Thomas Barrett, Jeremy Butterfield, Nicholas Teh, and James Weatherall for helpful discussions and comments. I also thank two anonymous referees for comments on the paper. I warmly thank the Black Hole Initiative at Harvard University for their hospitality during the winter of 2018-2019. I also thank my co-symposiasts at the 2018 PSA Biennial Meeting in Seattle, Thomas Barrett, Laurenz Hudetz, and James Weatherall, as well as the audience, for a lively and interesting session. I also thank the audience at the University of Chicago at Illinois. This work was supported by the Tarner scholarship in Philosophy of Science and History of Ideas, held at Trinity College, Cambridge, and by the Black Hole Initiative at Harvard University, which is funded by a grant from the John Templeton Foundation. 

\section*{References}
\addcontentsline{toc}{section}{References}

Arnold, V.~I.~(1989). {\it Mathematical Methods of Classical Mechanics}, Springer-Verlag, Second Edition.\\
\\
Barrett, T.~W.~(2018). `Equivalent and Inequivalent Formulations of Classical Mechanics'. Forthcoming in {\it The British Journal for the Philosophy of Science.} PhilSci: http://philsci-archive.pitt.edu/13092.\\
\\
Barrett, T.~W.~and Halvorson, H.~(2016). `Glymour and Quine on theoretical equivalence'. {\it Journal of Philosophical Logic}, 45 (5), pp.~467-483.\\
\\
Belot, G.~(1995). `Determinism and Ontology'. {\it International Studies in the Philosophy of Science}, 9 (1), pp.~85-101.\\
\\
Butterfield, J.~(2018). `On Dualities and Equivalences Between Physical Theories'. Forthcoming in {\it Space and Time after Quantum Gravity}, Huggett, N.~and W\"uthrich, C.~(Eds.).\\
\\
Carnap, R.~(1947). {\it Meaning and Necessity}. Chicago: The University of Chicago Press.\\
\\
Castellani, E.~and Rickles, D.~(2017). `Introduction to special issue on dualities'. {\it Studies in History and Philosophy of Modern Physics}, 59: pp.~1-5. doi.org/10.1016/j.shpsb.2016.10.004.\\
\\
Coffey, K.~(2014). `Theoretical Equivalence as Interpretative Equivalence'. {\it The British Journal for the Philosophy of Science}, 65, pp.~821-844.\\
\\
Curiel, E.~(2014). `Classical Mechanics is Lagrangian; It Is Not Hamiltonian'. {\it The British Journal for the Philosophy of Science}, 65 (2), pp.~269-321.\\
\\
De Haro, S.~(2015). `Dualities and emergent gravity: Gauge/gravity duality'. {\em Studies in History and Philosophy of Modern Physics}, 59, 2017, pp.~109-125. \\doi:~10.1016/j.shpsb.2015.08.004. PhilSci 11666.\\
\\
De Haro, S., Teh, N., Butterfield, J.N.~(2015). `Comparing dualities and gauge symmetries'. {\em Studies in History and Philosophy of Modern Physics}, 59, 2017, pp.~68-80. https://doi.org/10.1016/j.shpsb.2016.03.001\\
\\
De Haro, S.~(2016). `Spacetime and Physical Equivalence'. Forthcoming in {\it Space and Time after Quantum Gravity}, Huggett, N.~and W\"uthrich, C.~(Eds.). http://philsci-archive.pitt.edu/13243.\\
\\
De Haro, S.~and Butterfield, J.N.~(2017). `A Schema for Duality, Illustrated by Bosonization'. In: Kouneiher, J.~(Ed.), {\it Foundations of Mathematics and Physics one century after Hilbert}. Springer. http://philsci-archive.pitt.edu/13229.\\
\\
De Haro, S.~and Butterfield, J.N.~(2018). `On Symmetry and Duality'. This volume.\\
\\
De Haro, S.~(2018). `The Heuristic Function of Duality'. {\it Synthese}. https://doi.org/10.1007/s11229-018-1708-9\\
\\
De Haro, S.~(2018a). `Towards a Theory of Emergence for the Physical Sciences'. In preparation.\\
\\
De Haro, S.~(2019). `The Empirical Under-determination Argument against Scientific Realism for Dual Theories'. In preparation.\\
\\
De Haro, S.~and De Regt, H.~W.~(2018). `Interpreting Theories without a Spacetime'. Forthcoming in {\it European Journal for Philosophy of Science}.\\
\\
Dewar, N.~(2015). `Symmetries and the Philosophy of Language'. {\it Studies in History and Philosophy of Modern Physics}, 52, pp.~317-327.\\
\\
Dewar, N.~(2017). `Interpretation and Equivalence; or, Equivalence and Interpretation'. Forthcoming in: E.~Curiel and S.~Lutz (Eds.), {\it The Semantics of Theories}.\\
\\
Dewar, N.~(2018). `Sophistication about Symmetries'. {\it The British Journal for the Philosophy of Science}, forthcoming.\\
\\
Dieks, D., Dongen, J. van, Haro, S. de~(2014). `Emergence in Holographic Scenarios for Gravity'. 
{\it Studies in History and Philosophy of Modern Physics} 52 (B), 2015, pp.~203-216. doi:~10.1016/j.shpsb.2015.07.007.\\
\\
Fletcher, S.~C.~(2015). `Similarity, Topology, and Physical Significance in Relativity Theory'. {\it The British Journal for the Philosophy of Science}, 67 (2), pp.~365-389.\\
\\
Fraser, D.~(2017). `Formal and physical equivalence in two cases in contemporary quantum physics'. {\it Studies in History and Philosophy of Modern Physics}, 59, pp.~30-43. \\doi:~10.1016/j.shpsb.2015.07.005.\\
\\
Frege, G. (1892), `\"{U}ber Sinn und Bedeutung', {\em Zeitschrift f\"{u}r Philosophie und philosophische Kritik}, pp.~25-50; translated as `On Sense and reference', in P.T. Geach and M. Black eds. (1960), {\em Translations from the Philosophical Writings of Gottlob Frege}, Oxford: Blackwell.\\
\\
Frisch, M.~(2005). {\it Inconsistency, Asymmetry, and Non-Locality. A philosophical investigation of classical electrodynamics}. Oxford University Press.\\
\\
Glymour, C.~(1970). `Theoretical Equivalence and Theoretical Realism. PSA: Proceedings of the Biennial Meeting of the Philosophy of Science Association 1970, pp.~275-288.\\
\\
Glymour, C.~(1977). `The epistemology of geometry'. {\it No$\hat u$s}, pp.~227-251.\\
\\
Glymour, C.~(1980). `Theory and Evidence'. Princeton University Press, Princeton, NJ.\\
\\
Glymour, C.~(2013). `Theoretical Equivalence and the Semantic View of Theories'. {\it Philosophy of Science}, 80, pp.~286-297.\\
\\
Goldstein, H., Poole, C. and Safko, J.~(2002). {\it Classical Mechanics}, Addison Wesley, Third Edition.\\
\\
Halvorson, H.~(2012). `What Scientific Theories Could Not Be'. {\it Philosophy of Science}, 79, pp.~183-206.\\
\\
Halvorson, H.~(2013). `The Semantic View, If Plausible, Is Syntactic'. {\it Philosophy of Science}, 80, pp.~475-478.\\
\\
Halvorson, H.~and Tsementzis, D.~(2015). `Categories of Scientific Theories'. PhilSci: http://philsci-archive.pitt.edu/11923.\\
\\
Hodges, W.~(1997). `A Shorter Model Theory'. Cambridge: Cambridge University Press.\\
\\
Hudetz, L.~(2018). `Definable Categorical Equivalence'. PhilSci: http://philsci-archive.pitt.edu/14297.\\
\\
Huggett, N.~(2017). `Target space $\neq$ space'. {\em Studies in History and Philosophy of Modern Physics}, 59, 81-88. doi:10.1016/j.shpsb.2015.08.007.\\
\\
Lewis, D.~(1970). `General Semantics'. {\it Synthese}, 22, pp.~18–67. \\
\\
Lewis, D.~(1975). `Languages and Language'. In Keith Gunderson (ed.), {\em Minnesota Studies in the Philosophy of Science}, Volume VII, Minneapolis: University of Minnesota Press, pp.~3-35.\\
\\
Lewis, D.~(1983). `New Work for a Theory of Universals'. {\it Australasian Journal of Philosophy}, 61 (4), pp.~343-377.\\
\\
Lutz, S.~(2017). `What Was the Syntax-Semantics Debate in the Philosophy of Science About?' {\it Philosophy and Phenomenological Research}, XCV (2), pp.~319-352.\\
\\
Malament, D.~(2012). `Topics in the Foundations of General Relativity and Newtonian Gravitation Theory'. University of Chicago Press. \\
\\
M\o ller-Nielsen, T.~(2017). `Invariance, Interpretation, and Motivation'. {\it Philosophy of Science}, 84, pp.~1253-1264.\\
\\
Montague, R.~(1970). `Pragmatics and Intensional Logic'. {\it Synthese}, 22, pp.~68-94.\\
\\
Muller, F.~A.~(1997). `The Equivalence Myth of Quantum Mechanics---Part I'. {\it Studies in History and Philosophy of Modern Physics}, 28 (1), pp.~35-61.\\
\\
North, J.~(2009). `The `Structure' of Physics: A Case Study'. {\it The Journal of Philosophy}, 106, pp.~57-88.\\
\\
Prugove\v{c}ki, E.~(1981). `Quantum Mechanics in Hilbert Space', Second Edition. New York: Academic Press.\\
\\
Polchinski, J.~(2015). `Dualities of Fields and Strings'. {\it Studies in History and Philosophy of Modern Physics}, 59, 2017, pp.~6-20.\\
\\
Quine, W.~V.~O.~(1960). {\it Word and Object}. Cambridge, Massachusetts and London, England: The MIT Press. New edition, 2013.\\
\\
Quine, W.~V.~(1970). `On the Reasons for Indeterminacy of Translation'. {\it The Journal of Philosophy,} 67 (6), pp.~178-183.\\
\\
Quine, W.~V.~(1975). `On empirically equivalent systems of the world'. {\it Erkenntnis}, 9 (3), pp.~313-328.\\
\\
Read, J.~(2016). `The Interpretation of String-Theoretic Dualities. {\it Foundations of Physics}, 46, pp.~209-235.\\
\\
Read, J.~and M\o ller-Nielsen, T.~(2018). `Motivating Dualities'. Forthcoming in {\it Synthese}.\\
\\
Rickles, D.~(2017). `Dual theories: `same but different' or different but same'?' {\em Studies in History and Philosophy of Modern Physics}, 59, 62-67. doi:~10.1016/j.shpsb.2015.09.005.\\
\\
Rosenstock, S., Barrett, T.~W., Weatherall, J.~O.~(2015). `On Einstein Algebras and Relativistic Spacetimes'. {\it Studies in History and Philosophy of Modern Physics,} 52, pp.~309-316.\\
\\
Ruetsche, Laura (2011). {\it Interpreting Quantum Theories}. Oxford University Press.\\
\\
Suppe, F.~(2001). `Theory Identity'. In: {\it A Companion to the Philosophy of Science}, W.~H.~Newton-Smith (Ed.), pp.~525-527. Malden and Oxford: Blackwell.\\
\\
Teh, N.~J.~and Tsementzis, D.~(2017). ``Theoretical equivalence in classical mechanics and its relationship to duality'', {\it Studies in History and Philosophy of Modern Physics}, 59, pp.~44-54. doi.org/10.1016/j.shpsb.2016.02.002\\
\\
Trautman, A.~(1965). `Foundations and Current Problems of General Relativity'. In: {\it Lecutres on General Relativity}, Trautman, A., Pirani, F.~A.~E., Bondi, H.~(Eds), pp.~1-248. London: Prentice-Hall.\\
\\
Truesdell, C.~(1966). {\it The Elements of Continuum Mechanics}. Berlin Heidelberg New York: Springer-Verlag.\\
\\
van Fraassen, B.~C.~(1970). `On the Extension of Beth's Semantics of Physical Theories'. {\it Philosophy of Science}, 37 (3), pp.~325-339.\\
\\
van Fraassen, B.~C.~(1980). {\it The Scientific Image}. Oxford: Oxford University Press.\\
\\
van Fraassen, B.~C.~(2014). `One or Two Gentle Remarks about Hans Halvorson's Critique of the Semantic View'. {\it Philosophy of Science}, 81, pp.~276-283.\\
\\
Vickers, P.~(2013). {\it Understanding Inconsistent Science}. Oxford University Press.\\
\\
Von Neumann, J.~(1955). `Mathematical Foundations of Quantum Mechanics'. Princeton: Princeton University Press.\\
\\
Weatherall, J.~O.~(2015). `Categories and the Foundations of Classical Field Theories'. PhilSci: http://philsci-archive.pitt.edu/11587.\\
\\
Weatherall, J.~O.~(2016). `Are Newtonian gravitation and geometrized Newtonian gravitation theoretically equivalent?' {\it Erkenntnis}, 81 (5), pp.~1073-1091.\\
\\
Weatherall, J.~O.~(2016a). `Understanding Gauge'. {\it Philosophy of Science}, 83, pp.~1039-1049.\\
\\
Weatherall, J.~O.~(2019). `Equivalence and Duality in Electromagnetism'. arXiv:1906.09699.\\
\\
Wilson, M.~(2006). {\it Wandering Significance. An Essay on Conceptual Behavior}. Oxford: Oxford University Press.

\end{document}